\documentclass{article}
\usepackage[utf8]{inputenc}
\usepackage{comment}

\usepackage{capt-of}

\usepackage{geometry}
\usepackage[numbers]{natbib}
\usepackage{enumitem,amssymb}
\usepackage{aastex_hack}
\usepackage{graphics,graphicx}
\usepackage{fancyhdr}
\usepackage[svgnames]{xcolor}
\usepackage{floatrow}
\usepackage{wrapfig}
\usepackage{tcolorbox}
\usepackage[noblocks]{authblk}

\usepackage[hidelinks]{hyperref}
\hypersetup{
     colorlinks = true,
     linkcolor = teal,
     anchorcolor = blue,
     citecolor = teal,
     filecolor = blue,
     urlcolor = blue
     }

\newcommand\snowmass{\begin{center}\rule[-0.2in]{\hsize}{0.01in}\\\rule{\hsize}{0.01in}\\
\vskip 0.1in Submitted to the  Proceedings of the US Community Study\\ 
on the Future of Particle Physics (Snowmass 2021)\\ 
\rule{\hsize}{0.01in}\\\rule[+0.2in]{\hsize}{0.01in} \end{center}}

\begin{document}

\title{Snowmass 2021 Cosmic Frontier White Paper: \\ The Dense Matter Equation of State and QCD Phase Transitions}

\author[1]{Slavko Bogdanov} 
\affil[1]{Columbia Astrophysics Laboratory, Columbia University, 550 West 120th Street, New York, NY 10027, USA}

\author[2]{Emmanuel Fonseca} 
\affil[2]{Department of Physics and Astronomy, West Virginia University, White Hall, Box 6315, Morgantown, WV 26506, USA}

\author[3]{Rahul Kashyap}
\affil[3]{Institute for Gravitation and the Cosmos, Department of Physics, Penn State University, University Park, PA 16802, USA}

\author[4]{Aleksi Kurkela} 
\affil[4]{Faculty of Science and Technology, University of Stavanger, 4036 Stavanger, Norway}

\author[5]{James M.~Lattimer}
\affil[5]{Department of Physics and Astronomy, State University of New York at Stony Brook, Stony Brook, NY 11794, USA}

\author[6]{Jocelyn S.~Read} 
\affil[6]{Department of Physics, California State University, Fullerton, 800 N.~State College Blvd., Fullerton, CA 92831, USA}

\author[3,7]{Bangalore S.~Sathyaprakash} 
\affil[7]{Department of Astronomy and Astrophysics, Penn State University, University Park, PA 16802, USA}

\author[8]{H.~Thankful Cromartie}
\affil[8]{Cornell Center for Astrophysics and Planetary Science and Department of Astronomy, Cornell University, Ithaca, NY 14853, USA}

\author[9,10]{Tim Dietrich}
\affil[9]{Max Planck Institute for Gravitational Physics (Albert Einstein Institute), Am M\"uhlenberg 1, Potsdam 14476, Germany}
\affil[10]{Institute for Physics and Astronomy, University of Potsdam, Potsdam 14476, Germany}

\author[3]{Arnab Dhani}

\author[11,12]{Timothy Dolch}
\affil[11]{Department of Physics, Hillsdale College, 33 East College Street, Hillsdale, MI 49242, USA}
\affil[12]{Eureka Scientific, 2452 Delmer Street, Suite 100, Oakland, CA 94602-3017, USA}

\author[13,14]{Tyler Gorda}
\affil[13]{Technische Universit\"at Darmstadt, Department of Physics, 64289 Darmstadt, Germany}
\affil[14]{ExtreMe Matter Institute EMMI, GSI Helmholtzzentrum f\"ur Schwerionenforschung GmbH, 64291 Darmstadt, Germany}

\author[15,16]{Sebastien Guillot}
\affil[15]{IRAP, CNRS, 9 avenue du Colonel Roche, BP 44346, F-31028 Toulouse Cedex 4, France}
\affil[16]{Universit\'e de Toulouse, CNES, UPS-OMP, F-31028 Toulouse, France}

\author[17]{Wynn C.~G.~Ho}
\affil[17]{Department of Physics and Astronomy, Haverford College, 370 Lancaster Avenue, Haverford, PA 19041, USA}

\author[3]{Rachael Huxford}

\author[18,19]{Frederick K.~Lamb}
\affil[18]{Center for Theoretical Astrophysics and Department of Physics, University of Illinois at Urbana-Champaign, 1110 West Green Street, Urbana, IL 61801, USA}
\affil[19]{Department of Astronomy, University of Illinois at Urbana-Champaign, 1002 West Green Street, Urbana, IL 61801, USA}

\author[20]{Philippe Landry}
\affil[20]{Canadian Institute for Theoretical Astrophysics, University of Toronto, 60 St. George Street, Toronto, ON M5S 3H8 Canada}

\author[21]{Bradley W.~Meyers}
\affil[21]{Department of Physics and Astronomy, University of British Columbia, 6224 Agricultural Road, Vancouver, BC V6T 1Z1 Canada}

\author[22]{M.~Coleman Miller}
\affil[22]{Department of Astronomy and Joint Space-Science Institute, University of Maryland, College Park, MD 20742-2421, USA}

\author[1,23,24]{Joonas N\"attil\"a}
\affil[23]{Physics Department, Columbia University, 538 West 120th Street, New York, NY 10027, USA}
\affil[24]{Center for Computational Astrophysics, Flatiron Institute, 162 Fifth Avenue, New York, NY 10010, USA}

\author[25]{Risto Paatelainen}
\affil[25]{Department of Physics and Helsinki Institute of Physics, University of Helsinki, Helsinki, Finland}

\author[26]{Chanda Prescod-Weinstein}
\affil[26]{Department of Physics and Astronomy, University of New Hampshire, Durham, NH 03824, USA}

\author[25]{Saga S\"appi}

\author[21]{Ingrid H.~Stairs}

\author[27]{Nikolaos Stergioulas}
\affil[27]{Department of Physics, Aristotle University of Thessaloniki, 54124 Thessaloniki, Greece}

\author[28]{Ingo Tews}
\affil[28]{Theoretical Division, Los Alamos National Laboratory, Los Alamos, NM 87545, USA}

\author[25]{Aleksi Vuorinen}

\author[29,30]{Zorawar Wadiasingh}
\affil[29]{Astrophysics Science Division, NASA Goddard Space Flight Center, Greenbelt, MD 20771, USA} 
\affil[30]{Universities Space Research Association (USRA), Columbia, MD 21046, USA}

\author[31]{Anna L.~Watts}
\affil[31]{Anton Pannekoek Institute for Astronomy, University of Amsterdam, Science Park 904, 1090GE Amsterdam, The Netherlands}

\date{ }
\maketitle
\snowmass{}

\tableofcontents

\clearpage

\addcontentsline{toc}{section}{Executive Summary}
\begin{tcolorbox}
\section*{Executive Summary}
Our limited understanding of the physical properties of matter at ultra-high density ($\rho_s > 2.7 \times 10^{14}$\,g\,cm$^{-3}$ or $n_s = 0.16/$fm$^3$), high proton/neutron number asymmetry, and low temperature ($\lesssim 10^{10}~{\rm K}$) is presently one of the major outstanding problems in physics, owing to a number of challenges both in the experimental and theoretical realms \cite[see, e.g.,][for a review]{Watts16}. A plethora of well-motivated theoretical predictions for the state of matter in this regime have been proposed, ranging from normal nucleonic matter, to particle exotica such as hyperons, deconfined quarks, color superconducting phases, and Bose-Einstein condensates. \\ 

The principal macroscopic diagnostic of the state of dense matter is the pressure-density-temperature relation of bulk matter, the so-called equation of state (EOS). The EOS can, in principle, be employed to deduce crucial aspects of the microphysics, such as the role of many-body interactions at nuclear densities or the presence of deconfined quarks at high densities.  Matter in this extreme regime is known to only exist stably in the cores of neutron stars, the collapsed compact remnants of massive stars. As such, neutron stars can provide important insight into the properties of matter in a regime that is otherwise inaccessible.\\

In the upcoming decade it is possible to arrive at definitive answers to the following key questions:
    \begin{itemize} 
        \item \textbf{How does matter behave in the centers of neutron stars?} 
        \item \textbf{What are the physical properties of matter in the ultra-high density, large proton/neutron number asymmetry, and low temperature region of the quantum chromodynamics phase space?} 
 
    \end{itemize} 
These can be achieved through complementary measurements from electromagnetic (radio and X-ray) and gravitational wave astrophysical observations of neutron stars, combined with terrestrial laboratory constraints  and further theoretical investigations. 
This endeavor imposes the following requirements for facilities and resources in the upcoming decade and beyond: 
    \begin{itemize} 
        \item A next generation of gravitational wave detectors to uncover more double neutron star and neutron star-black hole mergers 
        \item Sensitive radio telescopes to find the most massive ($M>2$ $M_{\odot}$) and fastest spinning neutron stars  
        \item Large-area, high-time-resolution and/or high angular resolution X-ray telescopes to constrain the neutron star mass-radius relation 
        \item Suitable laboratory facilities for nuclear physics experiments to constrain the dense matter equation of state 
        \item Funding resources for theoretical studies of matter in this regime.  
        \item The availability of modern large-scale high performance computing infrastructure.\\
    \end{itemize}

The same facilities and resources would also enable significant advances in other high-profile fields of inquiry in modern physics such as the nature of dark matter, alternative theories of gravity, nucleon superfluidity and superconductivity, as well as a wide array of astrophysics, including but not limited to stellar evolution, nucleosynthesis, and primordial black holes. 
\end{tcolorbox}
\clearpage

\section{Introduction}

Neutron stars (NSs) are the densest objects in the Universe outside of black holes. Their density increases toward the center of the star reaching densities around 5--10$n_s$, several times the nuclear saturation density $n_s = 0.16/$fm$^3$. At these densities we currently do not know how matter behaves, what the phase structure is or what the dynamic degrees of freedom are. NSs offer a unique laboratory to study strongly interacting matter and the underlying theory of Quantum Chromodynamics (QCD) in the most extreme conditions. They have the potential to facilitate the discovery of novel exotic phases of matter in their cores including the appearance of strangeness in form hyperonic matter and ultimately the melting of nucleonic structure giving rise to novel form of cold quark matter. 

It is clear that understanding the structure of NSs and the state supranuclear matter is of immense importance. In this White Paper, we outline the various techniques that can be employed to this end, including electromagnetic (primarily radio and X-ray) observations of Galactic NSs and gravitational wave detections of merging NSs at cosmological distances, complementary laboratory measurements of neutron-rich nuclei, and additional theoretical investigations. 

The White Paper is organized as follows. In Section~\ref{sec:theory}, we summarize the theoretical underpinnings of NSs and the dense matter EOS. In Section~\ref{sec:lab} we describe terrestrial laboratory nuclear experiment approaches to constrain the nuclear symmetry energy. The measurements based on radio and X-ray observations of NSs are detailed in Sections~\ref{sec:radio} and \ref{sec:xray}, respectively, while in Section~\ref{sec:gw}, we describe the approach based on merger transient gravitational wave and correlated electromagnetic observations. We offer conclusions and recommendatations in Section~\ref{sec:conclusions}.

\section{Theoretical Equation of State }
\label{sec:theory}

The nature of matter at ultra-high densities ($\rho_s > 2.8 \times 10^{14}$\,g\,cm$^{-3}$), large proton/neutron number asymmetry, and low temperatures ($\lesssim 10^{10}~{\rm K}$) is, at present, one of the major outstanding problems in modern physics, owing to a number of challenges both in the experimental and theoretical realms, but a plethora of well-motivated theoretical predictions for the state of matter in this temperature-density regime have been proposed, ranging from normal nucleonic matter, to particle exotica such as hyperons, deconfined quarks, color superconducting phases, and Bose-Einstein condensates (for a review see~\cite{Watts16}). Matter in this extreme regime is known to only exist stably in the cores of NSs.

The structure of NSs is determined by the competition between self gravity and pressure of strong nuclear interactions keeping the star in a hydrostatic equilibrium. This interaction is described in the simplest case of non-rotating NSs by the Tolman-Oppenheimer-Volkoff (TOV) equations~\cite{Oppenheimer:1939ne, Tolman:1939jz}, which map the equation of state (EOS) of dense nuclear matter to the macroscopic properties of NSs, making the EOS the primary object of interest for the nuclear physics of NSs. A large community effort has been put to investigating the EOS using different \emph{ab-initio} calculations as well as various models.  The evidence from observations, in particular with the advent of multimessenger astronomy, is more complicated than the simple static systems described by the TOV equations and a significant push has been made in the past years to numerically solve the combined Einstein and relativistic-fluid-dynamic equations~\cite{Romatschke:2017ejr}.  

\begin{figure*}[t!]
\centering
\includegraphics[trim=0cm 1cm 0cm 0cm, clip,width=0.8\textwidth]{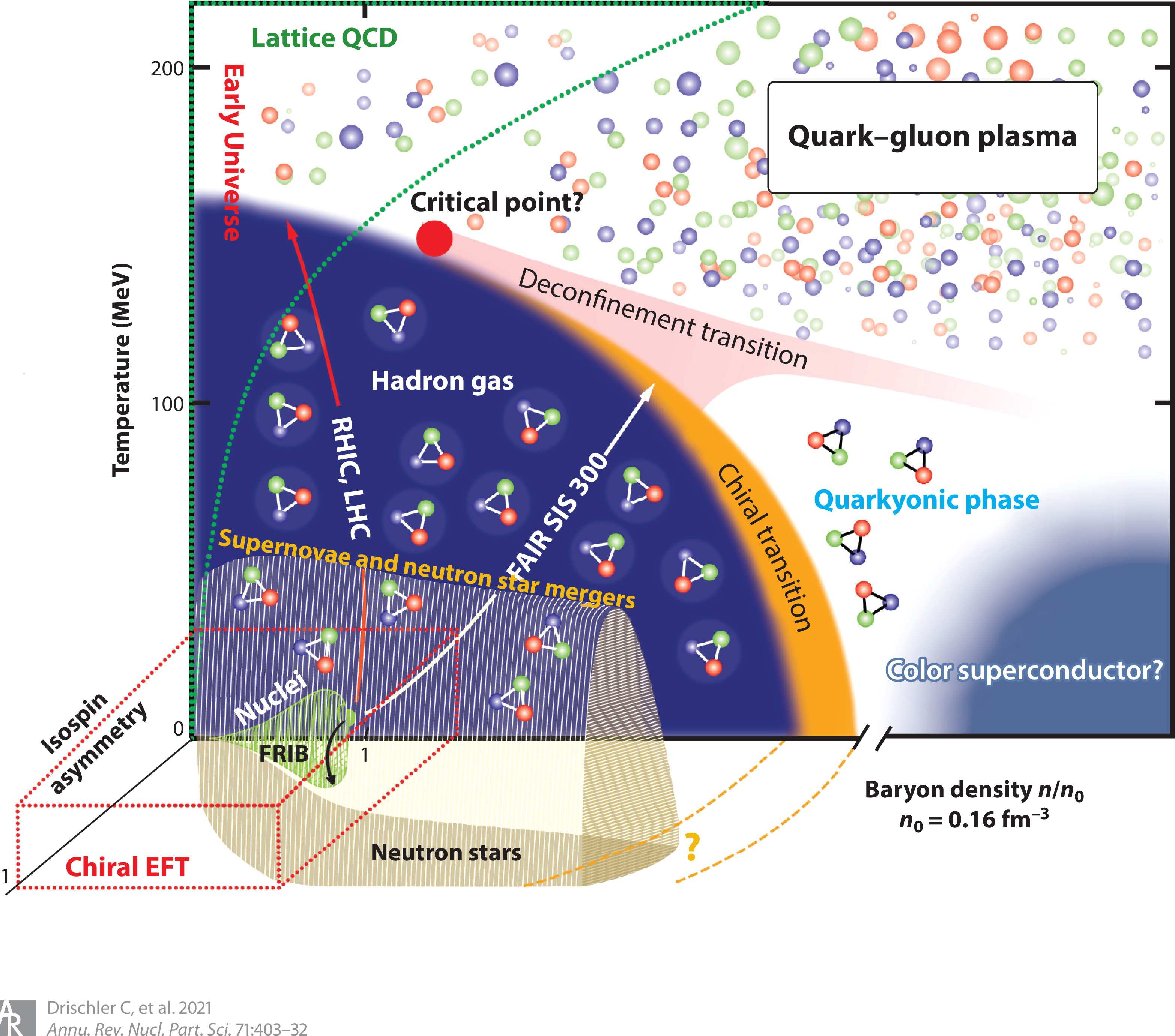}
\caption{Schematic view of the QCD phase diagram. The figure highlights regions probed by the RHIC, LHC, FAIR, and FRIB experiments, regions of validity for lattice QCD and chiral EFT, and environments reached in neutron stars, supernovae, and neutron star mergers. Abbreviations: EFT=effective field theory, QCD=quantum chromodynamics. Figure adapted from \cite{doi:10.1146/annurev-nucl-102419-041903}.}
\vspace{-0.3cm}
\label{fig:qcd_diagram}
\end{figure*}

Figure~\ref{fig:structure} displays a schematic figure of the NS structure. The crust and the outer core down to a depth of roughly $0.5$~km, where densities are of the order of nuclear density, is under good theoretical control~\cite{Baym:1971pw, Negele:1971vb}. Beyond that, our understanding of the structure relies on theoretical extrapolations. In particular, the phase of the inner core is currently unknown. Figure~\ref{fig:qcd_diagram} is a schematic view of the hypothesized phase diagram of QCD. It is a firm prediction of QCD wherein, at sufficiently high temperatures and/or densities, ordinary hadronic matter melts to a partonic form of matter---Quark Matter. 
\vskip5pt

\noindent
\begin{minipage}{0.48\textwidth}
In the regime of high temperatures and low baryon densities, the deconfiement transition to Quark Matter is well studied using lattice field theory~\cite{Fodor:2004nz,Aoki:2006we, Karsch:2001cy} and its existence is confirmed in two decades of experimentation with ultra-relativistic heavy-ion collisions~\cite{Margetis:2000sv} at the RHIC and the LHC. Further experimental program runs at the LHC aim to quantify the transport properties and the conditions of the onset of Quark Matter.

The deconfinement transition is a cross-over at low baryon densities but it has been long hypothesized that the transition becomes stronger with increasing baryon density. New theoretical arguments based on topological features of QCD have been recently put forward supporting the first-order-nature of the transition~\cite{Cherman:2018jir, Fukushima:2010bq}. The beam energy scan (BES) program at the RHIC~\cite{STAR:2017sal} and the future FAIR facility~\cite{CBM:2016kpk} are geared to discover the critical point separating the crossover transition from the first-order transition. The discovery of the critical point would have a profound impact on the physics of NSs.
Furthermore, heavy-ion collisions at beam energies up to 2 GeV per nucleon directly constrain the EOS at the density range of 1-2 $\rho_s$  \cite{Huth:2021bsp}.\end{minipage}
\hfill
\begin{minipage}{0.5\textwidth}
    \centering
    \includegraphics[width = 0.99\textwidth]{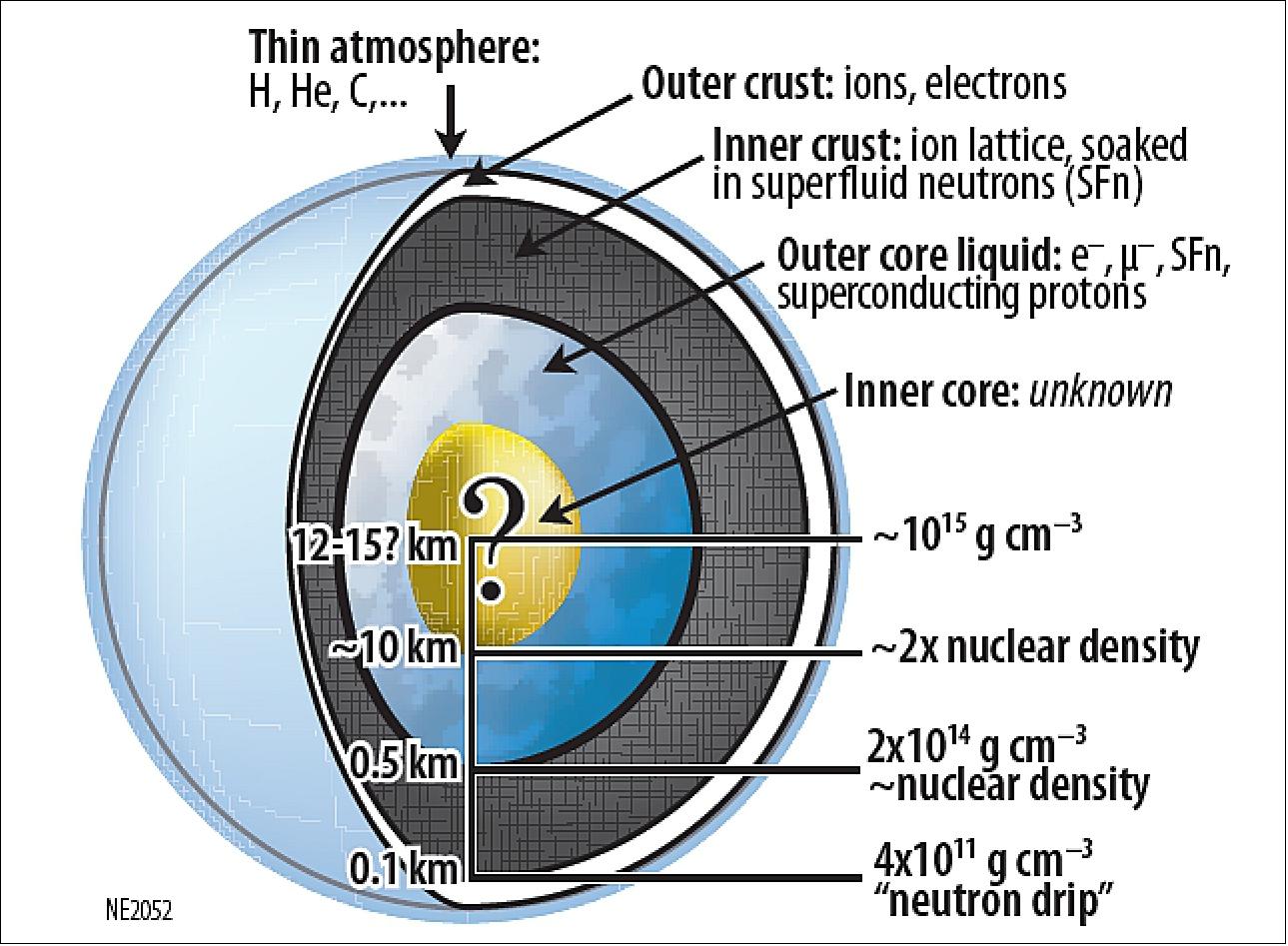}
    \captionof{figure}{A cartoon depiction of a neutron star and its internal structure. Credit: NASA/B.~Link. }
    \label{fig:structure}
\end{minipage}
\vskip5pt

At very-high densities, owing to an attractive interaction between quarks in QCD, it is expected that Quark Matter is in the form of a color superconductor, fundamentally affecting the transport properties of Quark Matter~\cite{Alford:2007xm, Alford:1997zt}. Based on large-$N_c$ arguments, it has also been speculated~\cite{McLerran:2018hbz} that, at low temperatures, there may be a further intermediate phase where the degrees of freedom inside the Fermi sea may be treated as quarks, while confining forces remain important near the Fermi surface.  Owing to the similarities with both the hadronic and Quark Matter phases, this hypothetical phase is dubbed the quarkyonic phase.

\subsection{Discovering Quark Matter Cores in Massive Neutron Stars}

The study of the QCD phase diagram presents formidable challenges in the limit of high baryon densities. On the theory side, lattice Monte-Carlo simulations are hampered by the Sign Problem, and observational information is similarly scarce due to ultrarelativistic heavy-ion experiments being limited to probing the hot and relatively baryon-dilute Quark Matter. The possible existence of the high-density Quark Matter inside massive NSs has been theorized a long time ago, but due to a combination of theoretical and observational challenges, this assertion has not been verified yet, but remains one of the grand challenges of nuclear astrophysics.
Recent years have witnessed unprecedented progress in the astrophysical observations of NSs, ranging from the famous observation of a gravitational-wave signal from a binary NS merger and its electromagnetic counterparts to more traditional X-ray and radio pulsar timing observations, providing insight into the macroscopic properties of NSs, such as their masses and radii \cite{Ozel_Freire_2016}, as well as superfluidity, superconductivity, thermal conductivity, and heat capacity of the matter in their interiors. This progress has opened a new window to ultra-dense matter and may even allow for a convincing detection of exotic phases of matter in NSs in the near future. However, in order to fully exploit the recent and upcoming observational data, a commensurate theoretical advances are also required.

In the absence of a smoking-gun observation that could only be explained by the presence of quark cores in NSs, a recent approach to the problem has been to combine all
available ab-initio theoretical results from nuclear and particle theory to robust observational information
on NSs in an effort to model-independently constrain the properties of matter in the centers of NSs of various masses.  The EOS reflects the phase diagram of the matter in the neutron star and a sufficiently accurate determination of the EOS may facilitate the identification of transition (a phase transition or a cross-over) from hadronic matter to deconfined Quark Matter. Certain features of the EOS differ qualitatively between the hadronic phase characterised by the confinement mass scale and the nearly conformal Quark Matter. Several such features have been suggested as indicators between the hadronic and Quark Matter phases, including the speed of sound, number of effective degrees of freedom, as well as the interaction measure \cite{Annala:2019puf, Kojo:2021ugu, Fujimoto:2022ohj}.  As shown in \cite{Annala:2019puf}, together with state-of-the-art observational data for the masses and tidal deformabilities of NSs, current results for the EOS of nuclear and quark matter
suffice to provide very stringent limits to how QCD matter needs to behave in order for quark cores to be absent even inside maximally massive NSs. There are several physical quantities, the constraining of which would play a major role in a future tightening of the current results, including:

\begin{itemize}
    \item EOS of nuclear matter at 1-2 nuclear saturation densities and of quark matter at 20-40 saturation densities,
    \item  Speed of sound of dense nuclear matter beyond saturation density
\item  Order of the deconfinement transition at small temperatures and high densities, and the magnitude of
the related latent heat,
\item  Radii and tidal deformabilities of NSs with accurately known masses,
\item  Resolved ringdown-phase observations of gravitational wave detections from NS merger events, and
\item  Electromagnetic observations of the NS merger remnants and the subsequent kilonova explosions.
\end{itemize}

In light of recent advances in the field of nuclear astrophysics, it is plausible that the eventual discovery of QM cores in massive NSs will not occur via a single dramatic observation, but rather through multiple theoretical and observational advances that work together in gradually eliminating the possibility of all NSs being composed of nuclear matter. Given the general importance of this question, improving the status of the aforementioned related results is a top priority in the field, and will hopefully receive considerable attention from all relevant research communities.
\subsection{Perturbative QCD}

While the Sign Problem prevents non-perturbative calculations of the EOS, at sufficiently high densities the EOS becomes accessible by \emph{ab-initio} calculations within perturbative QCD due the asymptotic freedom. The current state of the art in these calculations is the partial Next-to-Next-to-Next leading-order calculation at zero temperature, which becomes reliable at densities around and above $n=40 n_s$ \cite{Gorda:2021znl}. Effects of temperature and strange quark mass to the EoS have been discussed in \cite{Kurkela:2009gj, Kurkela:2016was,Gorda:2021gha}.  While the perturbative calculations become reliable only at densities which greatly exceed the densities which are reached within neutron stars, it has been recently demonstrated that the perturbative calculations constraint the EoS at neutron star densities \cite{Komoltsev:2021jzg}. These constraints arise from
the requirement that the EoS is causal, mechanically stable, and thermodynamically consistent at all densities, which, together with the known perturbative-QCD limit, imposes restrictions on the possible behaviour of the EoS at intermediate densities reached in NSs. Further improving the perturbative calculations will make these constraints more stringent. 

\subsection{Nucleon Superfluidity}
NSs are also critical testbeds for theoretical calculations of nucleon superfluidity since there are many observable behaviors of NSs that are understood to be due to the existence of superfluid neutrons and superconducting protons in the NS crust and core (see Figure~\ref{fig:structure}). For example, evolution of the spin rate of pulsars gives strong evidence for superfluid neutrons in the NS crust \cite{andersonitoh75,alparetal84,linketal99} and core \cite{hoandersson12}. Meanwhile, the observed temperatures of NSs of different ages require nucleon superfluids to be present in the core and even provide constraints on the critical temperature for neutron superfluidity $T_{\rm cn}(\rho)$ and proton superconductivity $T_{\rm cp}(\rho)$ \cite{gusakovetal04,pagetal04}. For example, the rapid cooling of the NS in the Cassiopeia~A supernova remnant necessitates a maximum $T_{\rm cn}=(5-10)\times 10^8$~K and $T_{\rm cp}>2\times 10^9$~K \cite{pagetal11,shterninetal11,shterninetal21}. Furthermore, type I versus type II superconductivity predicts decaying or long-lived magnetic fields in NS cores \cite{baymetal69} and thus has important implications for magnetic field evolution in NSs, from when NSs are young, such as in magnetars, until they are very old, like in millisecond pulsars. Therefore, continued astronomical observations to monitor NSs will provide improved understanding of properties of nucleon superfluidity. The interpretation of such observations would also benefit from consistent calculations of $T_{\rm cn}$ and $T_{\rm cp}$ from EOS models, such as those only just recently being done \cite{allardchamel22}, as well as effects such as entrainment \cite{chamel12,watanabepethick17}.

\subsection{Dark Matter in Neutron Stars and Exotic Compact Objects}
Despite extensive efforts using a wide array of direct and indirect detection methods, the particle nature of dark matter (DM) remains elusive. Because of their strong gravity and extreme core densities, NSs may accumulate DM from their environment by capturing DM particles after they scatter off of nucleons \cite{deLavallazFairbairn2010,KouvarisTinyakov2010}. 
Depending on the DM-nucleon interaction cross-section and the DM particle mass, the DM could form a core \cite{BertoneFairbairn2008,CiarcellutiSandin2011,GreshamZurek2019} or become admixed with the baryonic matter \cite{LeungChu2011,NelsonReddy2019}. 
DM could even be produced during the merger of two NSs~\cite{Dietrich:2019shr}. 
For certain DM models, a configuration of baryonic and non-baryonic matter is gravitationally stable \cite{PanotopoulosLopes2017,DeliyergiyevDelPopolo2019,QuddusPanotopoulos2020}. 
The DM concentration is then sensitive to the NS's age, temperature, mass and environment, producing diversity in the EOS-dependent observables, like tidal deformabilities, measured for the population. In other cases, DM may implode the stars \cite{Bramante:2017ulk}.
Alternatively, DM that is formed out of ultra-light scalar field could agglomerate on its own, forming compact objects that mimic true NSs and BHs~\cite{Cardoso:2017cfl, Sennett:2017etc, Clough:2018exo,Dietrich:2018jov}.
GW observations over the next decade, especially in the 3G detector era, combined with EM searches will give access to the full population of compact binary mergers, allowing us to search for ``smoking-gun'' EOS variability due to DM. In addition, X-ray observations of nearby NSs can be used to place constraints on the fraction of asymmetric DM inside a NS \cite{2022arXiv220803282R}. Therefore, observational studies of NSs using the techniques described below can, in principle, yield unique insight into the properties of DM.

\section{Connections with Nuclear Experiment and Theory}
\label{sec:lab}
Constraints on the EOS from mass and radius observations of NSs can be complemented with constraints on the nuclear symmetry energy obtained from nuclear experiments and ab-initio neutron matter theory.  It has been known for some time that there is a high degree of correlation between NS radii and the pressure of neutron star matter slightly above the nuclear saturation density.  The NS matter pressure at $n_s$ is nearly completely determined by the slope $L$ of the symmetry energy at the same density \cite{2001ApJ...550..426L}, which is experimentally probed by nuclear binding energies, dipole polarizabilities and neutron skin thicknesses of neutron-rich nuclei, and theoretically calculated from neutron matter studies.

The NS radius, in the absence of a strong first-order phase transition like that possibly associated with the QCD transition, is expected to vary by just a few tenths of a kilometer for NS masses in the range from $1.2M_\odot$ to $1.8M_\odot$, so uncertainties in a mass measurement are relatively unimportant.   Because the NS radius is most highly correlated with the pressure at about twice $n_s$, and $L$ is measured at $n_s$, other symmetry energy parameters such as $S_V$, $K_{sym}$ and $Q_{sym}$ (which are related to the NS matter energy and its second and third derivatives, respectively, at $n_s$) can also be relevant.  
The symmetry parameters are extractable, to varying degrees of precision, from the aforementioned nuclear experiments and theory.  As an example of the power of this correlation, the linear relation
$R/{\rm km}=(8.76\pm0.37)+(L/19.6{\rm~MeV})$ \cite{Zhao-Lattimer2022} characterizes to 68\% confidence about 200 of the most commonly used Skyrme-type nuclear interactions. It is likely that a more precise relation can be found when higher-order symmetry parameter ($K_{sym},Q_{sym})$ information is included. 
In general, nuclear data constrain linear relations among the symmetry parameters with more precision than their individual values, but different classes of experiments result in different correlations so that combining them can result in more accurate constraints in the individual parameters.  

Supporting current experimental estimates of the symmetry parameters are inferences from neutron matter theory.  Recently developed chiral effective field theory allows a systematic expansion of nuclear forces at low energies based on the symmetries of quantum chromodynamics (see \cite{doi:10.1146/annurev-nucl-102419-041903} for a review).  It exploits the gap between the pion mass (the pseudo-Goldstone boson of chiral symmetry-breaking) and the energy scale of short-range nuclear interactions established from experimental phase shifts.  It provides the only known consistent framework for estimating energy uncertainties.
These ab-initio results for pure neutron matter have the least systematic uncertainties, at present.  The predicted energy and pressure can be combined with the binding energy and saturation density of symmetric nuclear matter (well-determined from nuclear mass fits) to find values of $S_V$ and $L$ that are in remarkable agreement with existing information from fits to binding energies of thousands of stable nuclei. The framework has also recently been used to study the effect of finite temperatures to the EOS \cite{Keller:2020qhx, Keller:2022crb}. 

In practice, predictions of NS radii from nuclear mass fits and neutron matter theory are currently competitive with those from astrophysical observations.  Additional nuclear experiments, such as measurements of the skin thicknesses and dipole polarizabilities of nuclei, can also impact this problem.  Combined with binding energy fits and neutron matter theory, they presently constrain the symmetry parameters conservatively as $S_V=32\pm2$ MeV and $L=50\pm10$ MeV \cite{Lattimer21} (see also \cite{Essick:2021kjb}), suggesting $R=11.3\pm0.7$ km from the above linear relation.

Neutron skin and dipole polarizability measurements of neutron-rich nuclei are mostly sensitive to $L$, and until recently were largely consistent with expectations from binding energy fits and neutron matter theory.  But the small value of neutron skins, of order 0.1--0.2 fm, have resulted in large experimental uncertainties.  Therefore, new experimental constraints from parity-violating electron scattering measurements of the neutron skins of both Ca$^{48}$ and Pb$^{208}$ from the CREX \cite{2021APS..DNP.EA003P} and PREX \cite{2021PhRvL.126q2502A} experiments, respectively, have been particularly notable.  Such experiments have been argued \cite{Thiel:2019tkm} to have less systematic uncertainties than other types of measurements. Interestingly, the PREX experiment produced a measurement about 2 standard deviations larger than previous experimental results, while the CREX experiment produced a measurement smaller than, but within 1 standard deviation, of previous results.  Taken together, these two experiments are satisfied at the 90\% confidence level by $L$ values in the narrow range already predicted by binding energy fits and neutron matter theory, but the current large experimental uncertainties demand additional efforts to resolve this matter.  In addition, a new technique using FRIB and similar facilities to measure the proton radii of mirror nuclei such as Fe$^{54}$ and Ni$^{54}$ is an exciting development.  Proton radii can be measured with much higher precision than neutron radii, and their differences, in the absence of well-understood Coulomb effects, in mirror nuclei is equivalent to a neutron skin measurement \cite{PhysRevLett.85.5296}.

\section{Constraints from Radio Observations of Neutron Stars}
\label{sec:radio}

An important set of measurements in nuclear astrophysics comes from studying the timing properties of radio pulsars, rotating NSs that emit highly beamed radiation along their magnetic poles. An early example is the discovery and analysis of the ``Hulse-Taylor" radio pulsar-binary system that (indirectly) confirmed of the existence of gravitational radiation \cite{weisberg_gravitational_1981,taylor_new_1982}, and produced mass estimates with high precision. The recent discoveries of additional relativistic pulsar orbits and high-mass NSs, as well as the eventual detection of nanohertz-frequency gravitational radiation through pulsar timing, show that radio pulsars continue to serve as ideal laboratories for fundamental physics. 
    
\subsection{Measuring Neutron Star Masses in Pulsar Binary Systems} 

\begin{wrapfigure}{r}{0.48\textwidth}
    \centering
    \includegraphics[scale=0.59]{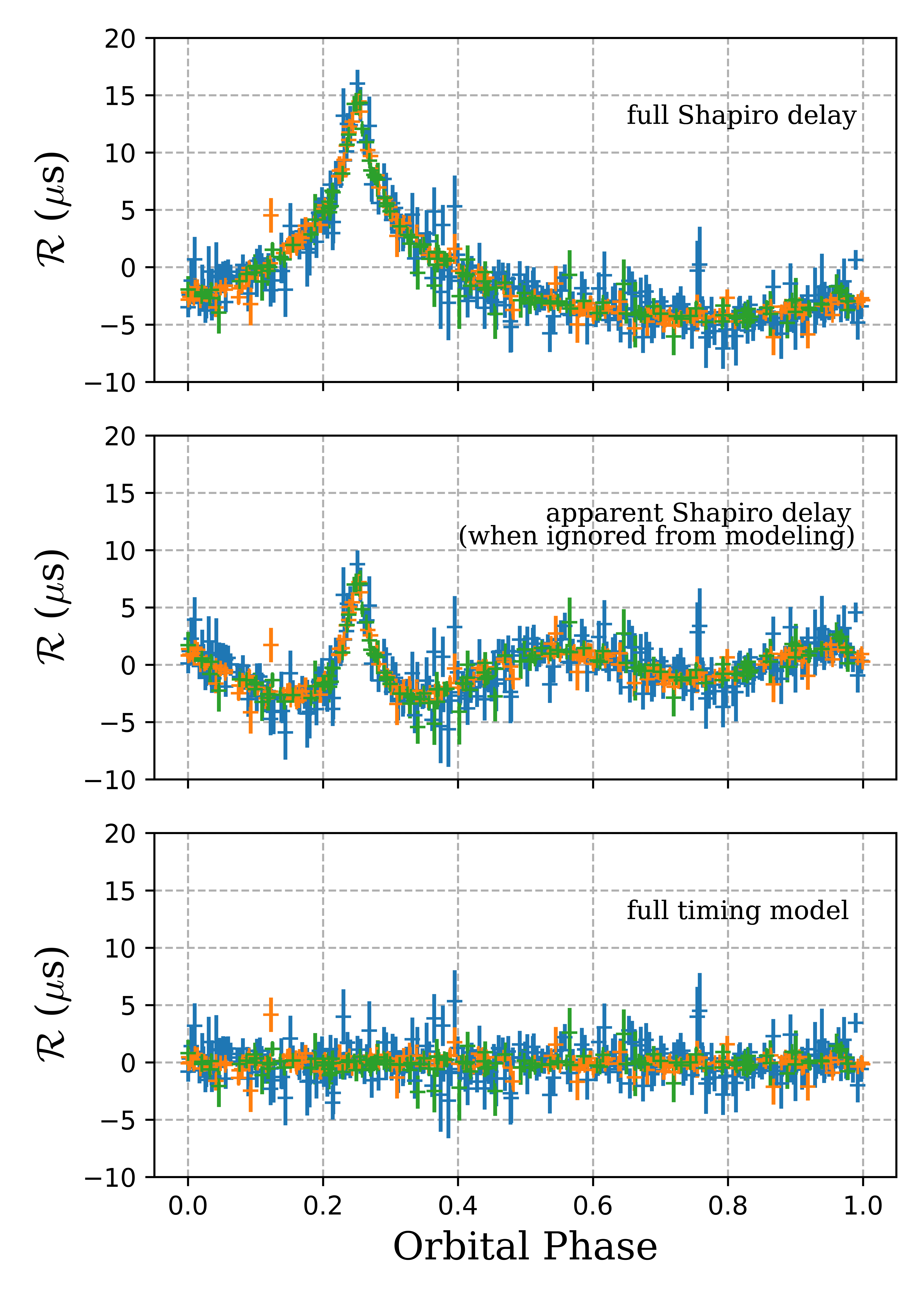}
    \caption{The signature of Shapiro delay in PSR J0740+6620 as a function of the pulsar's binary orbital phase, with colored points denoting pulse arrival-time residuals ($\mathcal{R}$), excess delay not accounted for by the timing model, measured at three different radio frequencies \citep[see][for details]{fcp+21}.  The top panel is the magnitude of the Shapiro delay if it is not included in the timing model but with all other model parameters fixed at their best-fit value. The middle panel shows the best-fit residuals obtained using an orbital model that does not account for general-relativistic
effects.  The bottom panel shows the fit with the full timing solution (including the relativistic Shapiro delay parameters).}
    \label{fig:shapiroJ0740}
    \vspace{-0.3cm}
\end{wrapfigure}    

One of the most effective ways of obtaining NS mass measurements through pulsar timing is by observing the relativistic Shapiro delay \cite{sha64} in binary orbits containing a pulsar. The Shapiro delay manifests as a small and periodic delay in pulse arrival times (of order $\sim$10\,$\mu$s) induced by the spacetime curvature of a pulsar's companion star. Measurement of the Shapiro delay directly yields the mass of the companion star as well as the geometry of the system; the NS mass is then determined by the Keplerian mass function. The Shapiro delay can be measured in a pulsar-binary system of any closed shape and size, so long as the companion mass and system orientation lead to delay amplitudes greater than $\sim$10 $\mu$s (see Figure~\ref{fig:shapiroJ0740}).

The Shapiro delay alone has provided some of the most impactful measurements of high-mass NSs to date \cite{dem10, cfr+20}. While rare, observations of $>$2\,M$_{\odot}$ NSs have greatly impacted our understanding of the NS interior EOS; for example, the initial measurement of the mass of PSR J1614$-$2230 effectively ruled out most contemporary non-baryonic EOS models (i.e., hyperons, kaons, Bose-Einstein condensates, free-quark stars, etc.), though recent work argues that the cores of maximally-massive NSs can be composed of quark-gluon matter under certain conditions (e.g., see \cite{Annala:2019puf}). Large NS masses also constrain the interaction between hadronic and strange-quark matter, and particular in phase transitions between the two states \cite{bhk+18}. On the other hand, measuring masses and radii of the lowest-mass NSs can constrain nuclear symmetry-energy parameters, at mean central densities close to the nuclear-saturation limit, that have strong implications for unknown aspects of supernova physics and other astrophysical phenomena \cite{sple05}.

Relativistic pulsar-binary systems, with orbital periods $\sim$ hours, are comparatively rare but nonetheless considered as idealized astrophysical environments due to the extreme gravitation in such systems. The discovery and long-term radio timing of merging double-neutron-star (DNS) systems, such as the Hulse-Taylor system, yield a variety of ``post-Keplerian" (PK) parameters -- the Shapiro delay being an example of PK effects -- that quantify corrections to purely Newtonian motion beyond O($|v|^2/c^2$), where $|v|$ represents the NS's orbital speed. An ensemble of PK parameters measured in a single pulsar-binary system can be used to test general relativity and directly estimate the binary-component masses \cite{ksm+06,fst10,wh16,kbv+20}.

A diverse set of tests for the presence of scalar fields as gravity mediators \cite{yunes_binary_2010,nair_improved_2020} are accessible when examining two/three-body systems contain pulsars with white-dwarf companions and of various sizes \cite{sfl+05,gsf+11}. These particular tests place stringent limits on tensor-scalar gravity theories, where notable effects include temporal variation of Newton's gravitational constant, violation of Lorentz invariance through excess acceleration of spinning, self-gravitating bodies within a preferred reference frame, and violations of equivalence principles through differing accelerations felt by bodies of different internal compositions and binding energies \cite{zdw+19}. The recent discovery and analysis of a pulsar in a three-body system with two white dwarfs \cite{rsa+14} have provided the strongest constraints of equivalence-violation principles to date \cite{agh+18,vcf+20}. Future discoveries and refined measurements will continue to provide exceptional strong-field tests of gravity.
\par
    
\subsection{Measurements of the Neutron Star Moment of Inertia}
With sufficiently long timing baselines, PK effects in DNS systems at O($|v|^4/c^4$) will become resolvable. An anticipated PK effect is apsidal motion due to spin-orbit interaction of a pulsar with a NS companion \cite{ds88}; this particular effect directly depends on the NS moment of inertia, which is itself a quantity that depends on both the NS mass and radius. The radio-timing measurement of the NS moment of inertia (e.g., \cite{hkw+20}) will provide a completely independent test for nuclear theory \cite{morrison_moment_2004,bejger_constraints_2005,fattoyev_sensitivity_2010,lim_predicting_2019}, being able to constrain the NS radius as well as nuclear-physics parameters at nuclear saturation density \cite{lattimer_constraining_2005,raithel_model-independent_2016,greif_equation_2020}, assuming general relativity is the correct theory of gravity. Such constraints may also be able to probe the possible existence of a hadron-quark phase transition at supranuclear densities, i.e., whether or not NSs may have a quark core \cite{alford_signatures_2019}.  An EOS-insensitive relation subject only to the conditions of causality and the existence of a $2M_\odot$ minimum to the NS maximum mass is $I=(0.87\pm0.05)(GM/Rc^2)^{1/2}MR^2$ \cite{Zhao-Lattimer2022}.  

In fact, the first meaningful constraint of an NS moment of inertia has already been reported \cite{Lattimer21} for the $1.338M_\odot$ primary of the double pulsar PSR J0737$-$3039, $I_{1.338}=64^{+65}_{-44}M_\odot$ km$^{2}$ after applying causality constraints.  This suggests, from the above EOS-insensitive relation, that $R_{1.338}=11.5^{+6.9}_{-6.3}$ km, consistent with expectations but not yet of the necessary accuracy to improve existing constraints.

The exquisite timing in PSR J0737-3039 and other DNS systems permitting resolution of PK effects also has important implications for current theories of strong-field gravitation \cite{ksm+06}. 
NS interiors are in both the strong-field and high spacetime-curvature regime where gravity modifications may appear in alternative theories and alter the interior structure. NS interiors thus provide both a strong-field and high-curvature test of gravity complementary to binary-pulsar tests. Given that the nuclear EOS is itself uncertain, this presents a possible observational degeneracy when one allows gravity to deviate from general relativity. A number of observable macroscopic NS properties obey quasi-universal relations, such as the I-Love-Q relation which is very weakly sensitive to the EOS \cite{yagi_i-love-q_2013}, allow for testing gravity with NSs in an EOS-independent manner, regardless of nuclear-physics uncertainties. The I-Love-Q relation is an order of magnitude more precise than the EOS-insensitive relation among I, M and R described above.  For example, combining independent measurements of the moment of inertia (from radio timing) and the tidal deformability (from gravitational waves) can rule out alternative gravity theories \cite{silva_astrophysical_2020} as well as test quasi-universality \cite{landry_constraints_2018}.

\section{Constraints from X-ray Observations of Neutron Stars}
\label{sec:xray}

\begin{figure}[t!]
\centering
\includegraphics[width=0.98\textwidth]{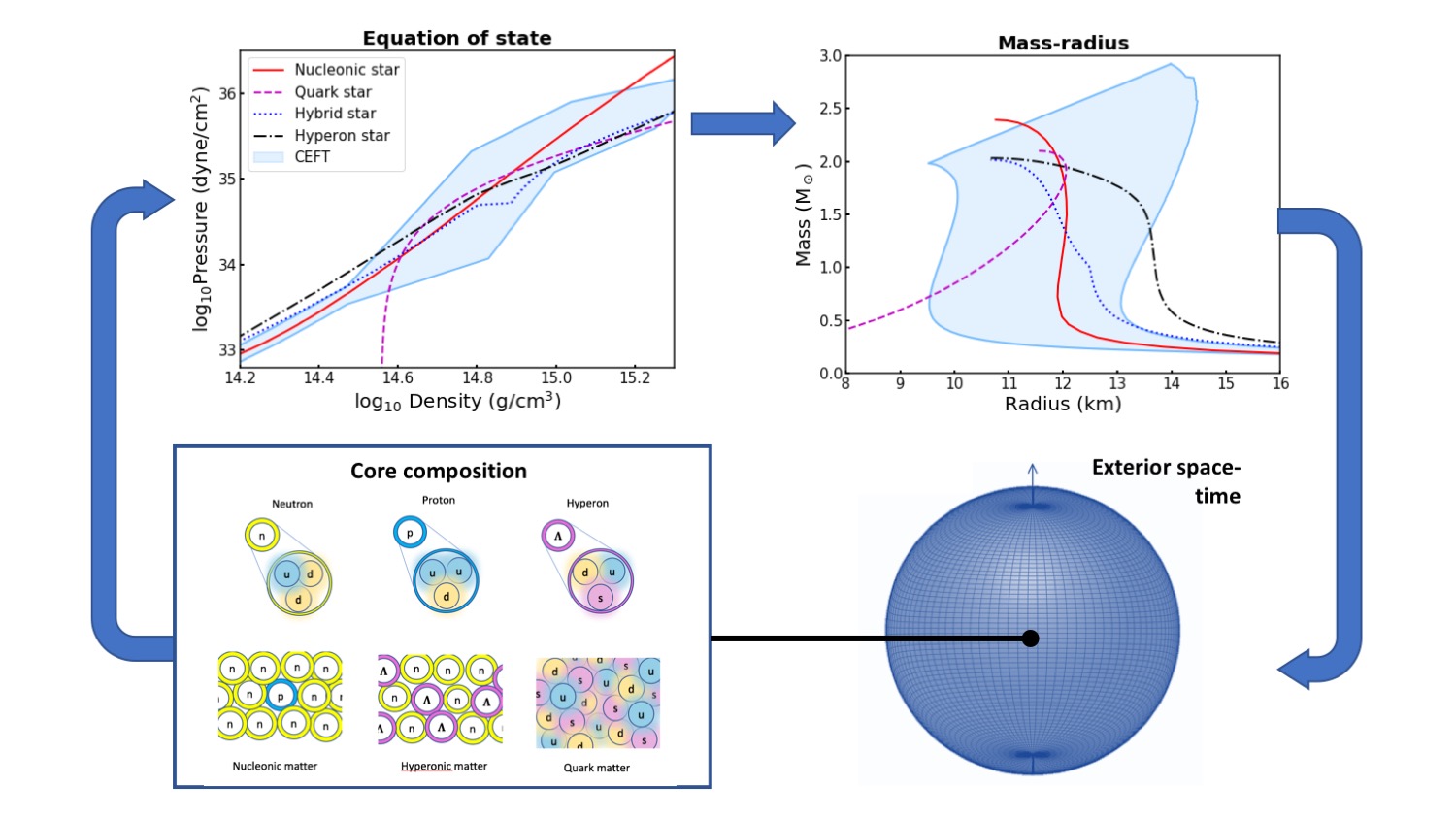}
\caption{The particle content and their interactions in the high density -- low temperature setting at the cores of NSs is highly uncertain.  Our lack of knowledge about these microphysical aspects (lower left:  uds = up down strange quarks) is encapsulated in the EOS (top left).  A sampling of plausible theoretical EOS models is shown that includes a nucleonic star (red) \cite{Akmal97}, a quark star (magenta) \cite{Li16}, a hybrid star consisting of a nucleonic outer core and quark matter inner core (blue) \cite{Zdunik13}, and a hyperon star with nucleonic outer core and hyperonic matter inner core (black) \cite{Bednarek_etal_2012}. The light blue region represents the approximate range spanned by the set of currently viable models \cite{hebeler13}.  The different EOS govern the global properties of the star such as $M$, $R$ and oblateness for a given rotation rate, via their influence on stellar structure (top right). These determine the exterior space-time properties of the star, which measurably alter the properties of the radiation propagating from the NS surface, encoding in it information about the EOS and the associated microphysics. Figure adapted from \cite{2019arXiv190303035R}.	}
\vspace{-0.3cm}
\label{fig:eosphysics}
\end{figure}

As noted previously, the mass-radius ($M$-$R$) relation of NSs has a strong dependence on the EOS of the dense matter in their interior  \cite[see, e.g.,][]{Lattimer01,Lattimer05,Ozel09,Read09a,hebeler13}. Thus, precise mass and radius measurements of several NSs using astrophysical observations can provide insight into the state of matter in their cores (Figure~\ref{fig:eosphysics}). 

Since it is not possible to directly sample the matter from the interior of NSs, it is necessary to rely on indirect inference using sensitive observations of their exteriors. This has motivated the development of a host of observational and data modeling techniques for inferring the mass and radius of NSs using electromagnetic observations primarily in the X-ray range ($\approx$0.1--30\,keV) of the surface thermal radiation from NSs. These approaches rely on the fact that the properties of the observed X-ray photons are greatly affected by the immense gravity in the vicinity of the star \cite[e.g.,][and references therein]{heinke13,miller13,potekhin14,bogdanov19}, which in turn, is determined by the stellar $M$ and $R$.

\subsection{X-ray Observational Techniques for EOS Constraints}

{\bf Pulse profile modeling.} For a spinning NS that emits X-rays from small regions on its surface (commonly referred to as ``hot spots''), its $R$ and $M$ can be estimated by modeling the observed rotation-induced X-ray pulsations. This is possible because the properties of the observed pulsed signal depend on $R$ and $M$ through the general and special relativistic effects on rotationally-modulated emission from the hot spots. Three varieties of NSs with surface hot spots suitable for such analyses are: rotation-powered millisecond pulsars, accretion-powered millisecond pulsars, and thermonuclear burst oscillation sources. The currently operating \textit{Neutron Star Interior Composition Explorer} (\textit{NICER}; see \cite{2016SPIE.9905E..1HG}) NASA X-ray timing mission is focusing on producing $R$ and $M$ measurement of \textit{a few} radio millisecond pulsars that produce thermal radiation \cite{Riley:2019yda,Miller:2019cac,miller21,riley21}. This pulse profile modeling technique (also known as waveform or light curve modeling) is mature, having been studied extensively over the last few decades \cite{Pechenick83,Miller98,Poutanen03,Poutanen06,Morsink07,Baubock13,Psaltis14,Algendy14,Nattila18}. Existing sophisticated models incorporate all practically important physics (atmospheric radiation properties, gravitational light-bending, Doppler boosting, aberration, propagation time delays, and the effects of rotationally-induced stellar oblateness) with a very high degree of accuracy. Interfacing these models with Baysian parameter estimation codes permits inference of either exterior space-time parameters or EOS parameters directly from the X-ray data \cite{Lo13,MillerLamb15,Riley18,Raaijmakers18,Raaijmakers:2021uju}.  Efforts for the {\em NICER} mission have produced inference codes highly optimized specifically for this purpose.

{\bf Extremely rapid rotators.} Discovering one or more NSs with spin rate $\gtrsim$1 kHz can lead strong and clean (i.e., systematics-free) EOS constraints. Each theoretical EOS has an associated prediction for the maximum spin rate before break-up (e.g, \cite{Haensel09}), so such NSs could, in principle, rule out a swath of proposed EoS models. X-ray pulse timing offers a particularly promising approach, since there is no evidence yet that the spin distribution of accreting millisecond pulsars drops off at high spin rates as seen in the radio pulsar sample \cite{Watts16}. Such rapidly spinning stars are likely to exhibit short-lived, intermittent X-ray pulsations, requiring hard X-ray detectors with high-time resolution and high instantaneous sensitivity.

{\bf Absorption line spectroscopy.} Detection of photospheric absorption lines from a NS, can, in principle, provide a measure of the stellar compactness $M/R$ \cite{Burbidge_1963} from their gravitational redshift.  In addition, precise measurement of the line shape may lead to constraints on $M$ and $R$ separately via line broadening physics \cite{Paerels_1997, Loeb_2003, Chang_etal_2005}. Absorption lines in NSs can arise from either atomic transitions or magnetic effects \cite[see, e.g.][for a review]{ozel13}.  Magnetic absorption features rely on an independent determination of magnetic field strength for redshift constraints. The most promising targets for atomic line features are thermonuclear (type-I) X-ray bursters, which generate bright episodic emission from the stellar surface, supply heavy elements to the photosphere through accretion, and are not affected by broad cyclotron lines owing to their weak magnetic field strengths \cite{Ozel_Psaltis_2003}.  

{\bf Radius expansion bursts.} Some thermonuclear bursts are powerful enough to exceed the Eddington critical luminosity, inflating the photosphere above the  stellar surface \cite{Tawara_etal_1984, Lewin_etal_1984}. For sources with measured distances (e.g., those situated in globular clusters), analysis of these photospheric radius expansion (PRE) allows measurements of $M$ and $R$ separately \cite[see, e.g.,][]{Ozel_etal_2010}. However, because PRE events exhibit a decline in blackbody temperature, most of the luminosity falls outside the X-ray energy coverage of most previous X-ray missions, resulting in large uncertainties in the measured Eddington luminosity \cite[e.g.,][]{Kuulkers_etal_2003, Guver_etal_2012}. In one approach, the Eddington flux is measured at the moment when the photosphere falls back onto the stellar surface as the burst cools \cite{Ozel_etal_2009, Steiner_etal_2010}.  In the other, the evolution of the spectrum of the  cooling tail of the burst is examined \cite{Nattila17}. The energy band of the current {\em NICER} mission, which extends as low as 0.25~keV, is well suited to such measurements.  Future missions with similar low-energy coverage but larger collecting area will be able to measure the photospheric evolution with greatly increased fidelity. Aside from determining the touchdown point more reliably, the evolution itself will provide additional determinations of $M$ and $R$, as both the Eddington limit and the gravitational redshift at each radius depend upon these parameters \cite{Damen_etal_1990}. 

{\bf Quiescent X-ray transients and cooling.} For a NS radiating uniformly from the entire surface, one can derive constraints on $M$ and $R$ from spectral fits to its X-ray emission if the temperature, composition of the topmost layer of the atmosphere, and its distance are known and the magnetic field is sufficiently weak ($\ll 10^{10}$\,G)  so as not to affect the opacity or temperature distribution on the NS surface. These criteria appear to be satisfied in quiescent low-mass X-ray binaries (qLMXBs) containing NSs, in particular, those located in globular clusters, to which the distances are well-determined \cite{Rutledge_etal_2002,Guillot_et_al_2013,Bogdanov_et_al_2016,Steiner_etal_2018}; plus additional targets in the field of the Galaxy with reliable parallaxes obtained with {\em GAIA}. Due to source crowding in the dense cores of globular clusters, effective studies of these targets require soft X-ray telescopes with high-angular resolution ($\le 1''$) imaging capabilities.

\section{Merger Transient Gravitational Wave and Electromagnetic Observations}
\label{sec:gw}

\noindent {{\bf Motivation}}. 

The Advanced LIGO~\cite{LIGOScientificCollaborationAasi2015} and Advanced Virgo~\cite{AcerneseAgathos2015} gravitational-wave (GW) detectors' discovery of a BNS merger in 2017, GW170817~\cite{AbEA2017b}, ushered in a new era of multimessenger astronomy.
Electromagnetic (EM) facilities observed a coincident short gamma-ray burst \cite{Monitor:2017mdv,Goldstein:2017mmi,AbEA2017e} and subsequent kilonova~\cite{AbEA2017f,Arcavi:2017xiz,Coulter:2017wya,Lipunov:2017dwd,
Soares-Santos:2017lru,Tanvir:2017pws,Valenti:2017ngx}, demonstrated that BNSs are prolific formation sites for many of the heavy elements found in nature~\cite{KaKa2019,Watson:2019xjv}. 
through the coordinated action of $\sim$\,70 
Late-time near-infrared observations showed BNS mergers to be sites of the rapid neutron-capture process (r-process), i.e. nucleosynthesis producing many of the heavy elements found in nature~\cite{KaKa2019,Watson:2019xjv}. 
Analyses of the GW data provided limits on the size of NSs
~\cite{TheLIGOScientificCollaboration2018, Annala:2017llu, Bauswein:2017vtn,De2018,Most2018,Fattoyev2018,Radice2018a,2019EPJA...55...97T,2020NatAs.tmp...42C,2020PhRvD.101l3007L, Dietrich:2020lps},
and joint GW/EM analyses enabled an independent measurement of cosmological parameters~\cite{Abbott:2017xzu,Kashyap:2019ypm,Coughlin:2019vtv,Coughlin:2020ozl,Dietrich:2020lps}.
The most recent LIGO-Virgo observing run produced another BNS merger \cite{lvc2020} as well as two NS-BH signals \cite{lvc2020}.

\begin{figure}[t!]
\centering
\includegraphics[width=0.98\textwidth]{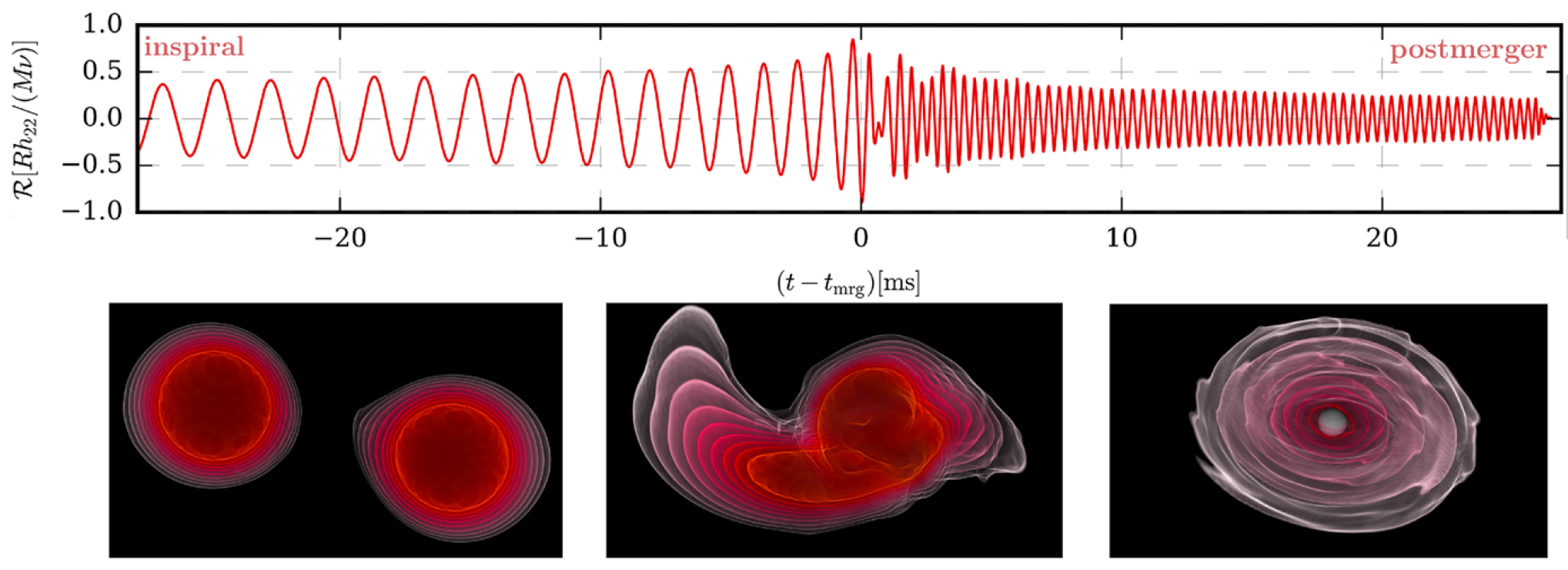}
\caption{Numerical relativity simulation of a binary NS merger showing the GW signal and the matter evolution. Top panel: The gravitational signal emitted during the last orbits before the merger (late-inspiral phase) and during the postmerger phase of the BNS coalescence. Bottom panels: Rest-mass density evolution for the inspiral (left panel), the merger (middle panel) and the postmerger phase after the formation of the black hole (right panel). Figure adapted from \cite{2020arXiv200402527D}.}
\label{fig:bns_merger}
\end{figure}

The current network of LIGO, Virgo, and KAGRA~\cite{AsoMichimura2013} will continue commissioning and observing runs through to c.~2028. 
The network expects to reach design sensitivity in the coming observing runs, at which up to tens of BNS mergers per year may be detected. 
Planned upgrades in c.~2025, including the addition of a fifth observatory (LIGO-India~\cite{IYER-LIGOINDIA}), will extend their reach for binary NS detections from ${\sim}100$\,Mpc (in 2020) to ${\sim}300$\,Mpc \cite{Aasi:2013wya-arxiv}. 
Third generation (3G) observatories, such as the Einstein Telescope \cite{Punturo:2010zz,Hild:2010id} and Cosmic Explorer \cite{Martynov2019,Reitze:2019iox}, currently in the design stage, may observe BNSs and BHNSs out to the era of peak star formation in our Universe.  Proposed dedicated high-frequency GW detectors \cite{Martynov2019, Ackley:2020atn} may target the merger and post-merger phases of NS coalescence.

At the same time, EM surveys such as the Vera C.\ Rubin Observatory's (VRO) Legacy Survey of Space and Time (LSST)~\cite{Ivezic:2008fe} and the Dark Energy Spectroscopic Instrument (DESI)~\cite{Aghamousa:2016zmz}, as well as the advent of 30\,m class telescopes, will push the boundaries of depth and cadence.
The combined operation of GW and EM observatories in the next decade will provide a unique opportunity to resolve long-standing questions such as the nature of the NS EOS including possible {\it phase transitions} to exotic matter, an independent measurement of the {\it expansion rate of the Universe}, and a possibility to reveal the nature of {\it dark matter} (DM). There is also the prospect of detecting long-duration GWs from single NSs; such GWs could be produced by a quadrupole
moment or fluid oscillation and would provide unique insights into
dense nuclear matter such as its elasticity and transport properties
and exotic phases like hyperons and quarks \cite{2018ASSL..457..673G,2022arXiv220606447R}

\noindent {{\bf Dense matter in NS interiors}}. 
GW and EM observations constraining NS properties are sensitive to the dense matter EOS due to the strong connections between astrophysical observables and microphysical interactions. The presence of a NS imprints on to gravitational-wave inspiral signal information about cold, dense matter up to a few times nuclear saturation density. The post-merger gravitational wave signal, on the other hand, contains information about hot dense matter EOS at even greater densities. 
Analyses of GW170817 already provide stringent constraints of the radius of a typical $1.4 M_{\odot}$ NS
~\cite{TheLIGOScientificCollaboration2018,Bauswein:2017vtn, De2018,Most2018,Fattoyev2018, Radice2018a, 2019EPJA...55...97T,2020NatAs.tmp...42C, 2020PhRvD.101l3007L, raaijmakers20, Dietrich:2020lps}. 
Several of these estimates  included multi-messenger input, combining GW and EM observations of BNS mergers, radio observations of massive pulsars~\cite{Antoniadis:2013pzd,Arzoumanian:2017puf,cfr+20}, X-ray observations by NASA's Neutron Star Interior Composition Explorer (NICER) mission~\cite{Miller:2019cac, Riley:2019yda}, and nuclear-physics constraints~\cite{Hebeler:2009iv, 2019EPJA...55...97T} (see also, e.g., \cite{2019PrPNP.10903714B, 2019JPhG...46l3002G, 2019EPJA...55...80R, Essick:2019ldf, 2020arXiv200203863R, 2020arXiv200402527D, 2020arXiv200406419B, 2020arXiv200603168C, 2020arXiv200407744E}).

\begin{figure}[th!]
    \centering
    \includegraphics[width=0.50\textwidth]{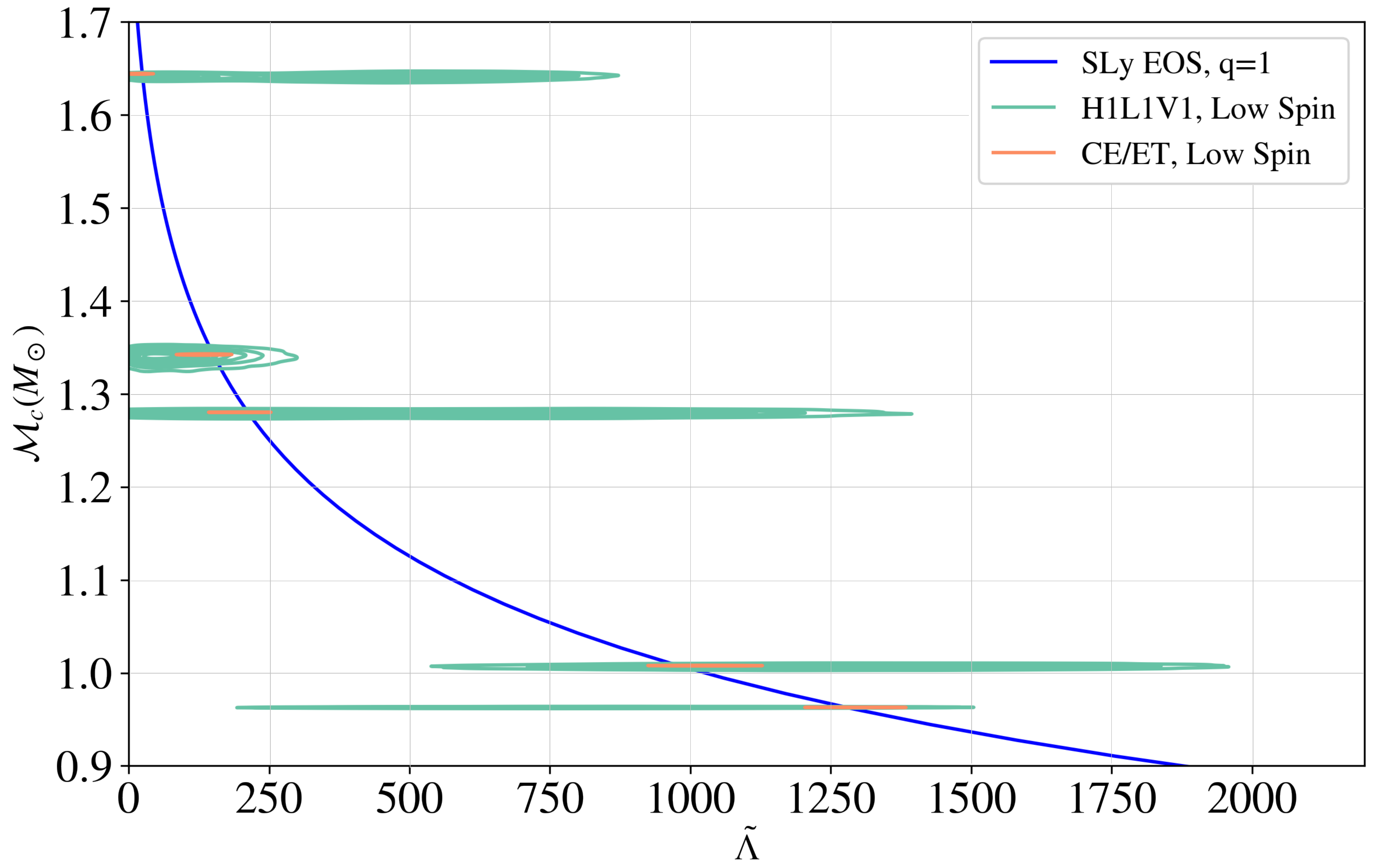}
    \includegraphics[width = 0.49\textwidth]{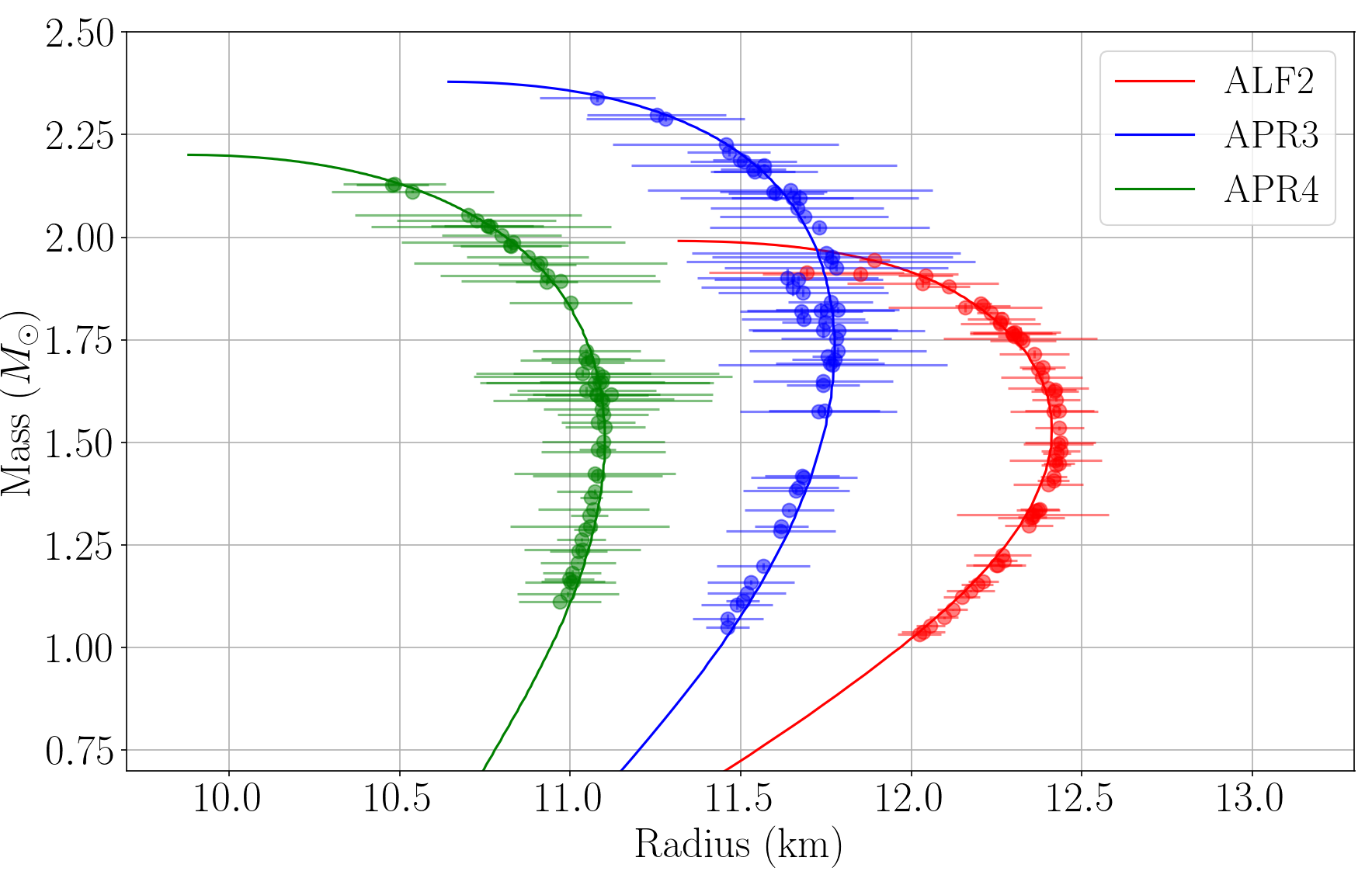}
    
    \caption{\textit{Left:} Simulated GW observations of neutron-star mergers of 5 moderately loud signals (signal to noise ratio $>13$) with \textit{LIGO} and \textit{Virgo} (H1L1V1), randomly drawn from component masses between 1 and 2 $M_\odot$. Merger waveform dependence produce posterior distributions encompassing 90\% probability around the assumed $M$-$\Lambda$ relation (blue line), shown here with mass ratio ($q$) dependence suppressed. The recovery is compared to the same signals recovered using \textit{Cosmic Explorer} and the \textit{Einstein Telescope} (CE/ET). CE/ET would also detect many additional signals in the same time period. Details will appear in \citep{WuchnerPrep}. \textit{Right:} Mass-radius curves for three different equations of state along with the accuracy with which neutron star radii could be measured for 30 loudest events using Fisher matrix approximation \cite{Borhanian:2020ypi} in a network consisting of two Cosmic Explorer detectors and an Einstein Telescope~\cite{Kashyap:2022wzr, Huxford:2022ab}. Errors in mass measurement are much smaller, hence are not visible in the plot.}
    \label{fig:XGradcorr}
\end{figure}

In the next decade, stronger constraints from combining information from multiple events~\cite{2020PhRvD.101l3007L, Bose2018, Yang2018} will enable percent-level measurements of the NS radius and the fundamental interactions of dense neutron-rich matter~\cite{2013PhRvL.111g1101D,2015PhRvD..92j4008C,2015PhRvD..91d3002L,2019PhRvD.100j3009H,2020arXiv200402527D,Bauswein2012,Bauswein2012a,bsj14,Clark2014,Clark2016,Torres-Rivas2019,Breschi:2019srl,2020PhRvD.101h4039V,Easter:2020ifj, 2020arXiv200407744E,Montana:2018bkb}. 
At this stage, systematic uncertainties due to modeling, which are currently negligible for individual events, are likely to lead to  biases in the final result. In the era of CE-ET, these systematics will be larger than the statistical errors coming from the noise in the detectors. Hence, modeling the systematics bias is required to constrain the precise model EOS \cite{Kashyap:2022wzr}.
On the left panel, Fig.~\ref{fig:XGradcorr}, improvements in the measurement of chirp mass and effective tidal deformability are shown for five events. Such improvements will lead to highly accurate measurements of NS radii. In the right panel of Fig.~\ref{fig:XGradcorr}, we show 30 simulated events for three EOSs (ALF2, APR3 and APR4) with highest SNR observed in the detector combination with two cosmic explorers and one Einstein Telescope. Both radius and mass errors are shown for each circle which depend on the EOS used for the simulation. 

It will be increasingly important to control systematic errors from prior assumptions about the EOS and the mass distribution of NSs in merging binaries~\cite{Wysocki:2020myz}, emphasizing the need to understand the limitations of EOS models and the population of sources as a whole. EOS-insensitive (quasi-universal) relations can be tested with GW observations~\cite{Samajdar:2020xrd} and applied to improve the accuracy of source properties~\cite{2019PhRvD..99h3016C}.  Nuclear physics may imprint on the GW source mass distribution itself~\cite{Farmer:2020xne}. Precise knowledge of the EOS or the mass distribution, particularly any sharp features therein, may enable GW observations alone to constrain the local expansion rate of the universe~\cite{Messenger:2013fya,
PhysRevD.86.023502, Farr_2019, PhysRevD.95.043502} and directly calibrate standard candles such as type Ia SNe \cite{Gupta:2019okl}.
Going beyond the equilibrium EOS, future BNS signals may reveal dynamical processes, e.g. stellar oscillations, that probe transport properties of normal matter or exotica such as boson condensates, hyperons, and quarks~\cite{2018ASSL..457..575H,2018ASSL..457..673G,2018RSPTA.37670285H,2019JPhG...46l3002G}. 

\noindent {{\bf New regions of the QCD phase space, black hole formation.}}
Both density and temperature increase dramatically after two NSs collide. Therefore, observing merging NSs enables us to probe not only the low-temperature QCD phase diagram, but provides a window on the most extreme conditions in the universe, which will shed additional light on the existence of phase transitions or exotic forms of matter.
While these transitions or the breakdown of purely hadronic models could be detected during the inspiral phase under certain conditions~\cite{2020PhRvD.101d4019C, Blacker:2020nlq, Pang:2020ilf, 2020arXiv200407744E}, their existence may be more apparent during the post-merger phase when densities are higher~\cite{2017ApJ...842L..10R,2019PhRvL.122f1102B,2019ApJ...881..122D,2019Univ....5..156H,2019PhRvL.122f1101M,2019arXiv191209340W,2020EPJA...56...59M}.
The required sensitivity to detect the post-merger GW signal from a GW170817-like event is likely to be achieved with future GW detector upgrades~\cite{Clark2016,Torres-Rivas2019}, and steadier rates are expected with next-generation facilities\cite{Srivastava:2022slt}.
Stacking data from multiple distant events may first reveal the post-merger \cite{Bose2018,Yang2018}. 
Increasing detector sensitivity allows us to observe post-merger GW signatures and determining the threshold mass for prompt collapse of the merger remnant to a BH will provide stringent constraints on the EOS and verify possible strong phase transitions in the NS EOS~\cite{2020PhRvD.101d4006A,Bauswein:2013jpa,Bauswein:2017vtn,2020arXiv200400846B,Kashyap:2021wzs}. 
In addition, knowledge of the exact threshold mass can inform future EM counterpart searches, as systems undergoing prompt collapse are most often not connected to bright, detectable EM signals, e.g.,~\cite{Coughlin:2020fwx,Coughlin:2019zqi}, and can be used to distinguish between different types of compact binary progenitors~\cite{2020arXiv200701372E}, possibly resolving long-standing issues associated with the lower ``mass gap'' between observed NSs and BHs, as well as uncertainties in the core-collapse supernova (CCSN) explosion mechanism~(e.g., \cite{Fryer_2001, Belczynski_2012}) and the nucleosythesis therein. 

\begin{figure*}
    \centering
    \includegraphics[width = 0.6\textwidth]{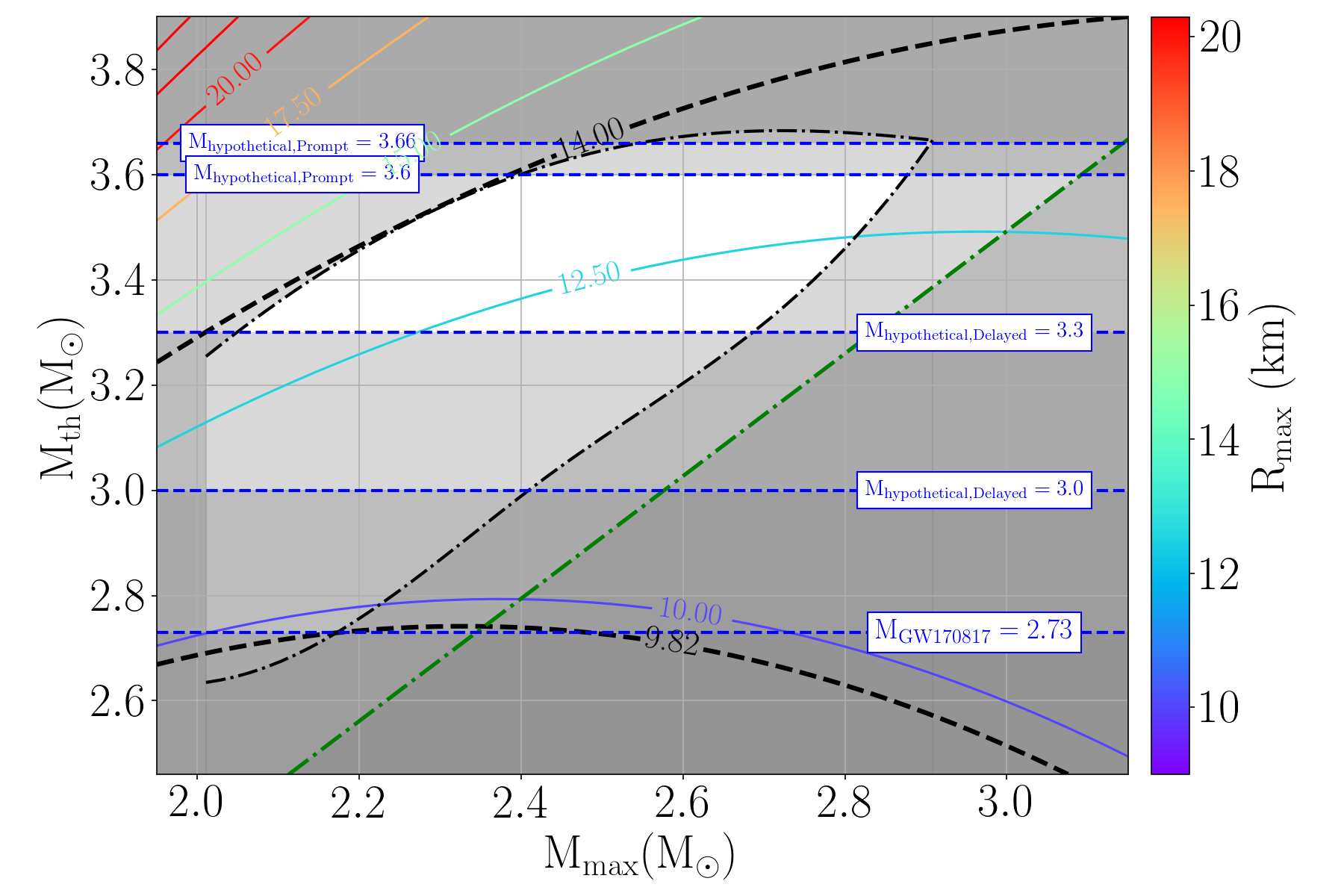}
    \caption{Constraints from GW170817 along with a few hypothetical scenarios expected in next generation GW detectors for constraining upper and lower limit of maximum mass and radius of maximum mass NS. Similar plots exist for radii of 1.4 M$_\odot$ and 1.6 M$_\odot$ NSs.  Figure adapted from \cite{Kashyap:2021wzs}. }
    \label{fig:pbhmmax}
\end{figure*}

Additionally, the precise value of the threshold mass to prompt collapse is found to be proportional to the maximum mass that a particular EOS supports \cite{Shibata:2005ss,Hotokezaka:2011dh}. The correlations found in numerical relativity simulations are useful in constraining the properties of EOS such as radius of 1.4 M$_\odot$, 1.6 M$_\odot$ and maximum mass NS \cite{Bauswein:2013jpa,Bauswein:2017vtn,Koppel:2019pys,Bauswein:2020xlt,Tootle:2021umi}. New correlations have been found using a large collection of EOSs, where classification of events as either prompt or delayed collapse will also help us in placing upper and lower limits on the maximum mass of NSs as well as the radius of 1.4 $M_\odot$, 1.6 $M_\odot$ and maximum-mass NSs \cite{Kashyap:2021wzs} (see Fig.~\ref{fig:pbhmmax} for GW170817 and a few hypothetical BNS merger events with their assumed outcomes). Most importantly, accurate modeling of tidal effects during the inspiral phase as well as greater coverage of the parameter space (e.g., total mass, mass ratio and longer evolution time) for accurate post-merger will put heavy demand on computational resources. give upper limits on the radius and maximum mass of neutron stars but also revealed new constraints on radius and tidal deformability of a 1.4 M$_\odot$ NS \cite{Kashyap:2021wzs}.

 Post-merger signals from BNS mergers will become visible in 3G detectors. In addition to the properties of hot EOS, QCD phase transition leaves subtle imprints on the post-merger signal such as a shift in the frequency of peak frequency $f_2$ in the post-merger gravitational waves \cite{Blacker:2020nlq}. New correlations have been observed between $f_2$, effective tidal deformability $\tilde{\Lambda}$ and symmetric mass ratio $\eta$ \cite{Bernuzzi:2015rla}. QCD phase transition could break such correlations and hence an opportunity of its detection arises in BNS mergers that undergo QCD phase transition. However, to infer the full nature of the post-merger signal, one must develop an accurate waveform model that could be used in Bayesian inference. Several attempts have been made to develop such a model \cite{Clark2016, Chatziioannou:2017ixj, Tsang:2019esi, Easter:2020ifj, Soultanis:2021oia}. More recently, new post-merger waveform models have been developed using numerical relativity simulations in the parameter space of total mass, $M \in [2.4,3.4] M_\odot$, mass ratio, $q < 2$ and non-precessing dimensionless spin up to 0.2, that have faithfulness high enough to detect the phase transition in ET \cite{Breschi:2019srl,Breschi:2022xnc,Breschi:2022ens}. 

\noindent {{\bf Matter outflows and r-process nucleosynthesis.}}
Postmerger GW and EM observations trace the dynamic aftermath in the extreme environments created by compact binary mergers.
Folding in EM observations provides an opportunity to independently constrain the source properties and to obtain a better understanding of the physical processes and outcomes of mergers~\cite{CoDi2018b,Dietrich:2020lps}. 
The successful search and detection of the \emph{kilonova} associated with GW170817 revealed NS mergers as a critical site of rapid neutron-capture process (r-process) nucleosynthesis.
Combined with the inferred BNS merger rate~\cite{Aasi:2013wya,lvc2020} and detailed nucleosynthesis simulations, future multimessenger observations of NS mergers will elucidate the processes that make the heavy elements~\cite{Lattimer:1974slx,AbEA2017g}.
An interdisciplinary effort implicating GW and EM observations, numerical relativity (NR) simulations, and nuclear theory calculations will enable detailed predictions of the abundances of individual elements.
In the future, we expect that wide field-of-view, optical survey instruments are most likely to find fast-fading kilonovae. While today's searches for these counterparts are difficult due to their large sky localizations~\cite{Fairhurst:2009tc,Singer:2014qca,Essick:2014wwa}, 3G detectors will significantly improve the localizability. Information from the inspiral GWs together with NR simulations can inform EM follow-up campaigns by making robust predictions about potential matter outflows and light curves~\cite{Coughlin:2019zqi,BarbieriSalafia2019}. Together,  GW and EM observations may reveal exotic components like primordial black holes~\cite{2020PhRvD.101j3008C,2020PhRvD.102b3025F,Hinderer:2018pei,Fuller:2017uyd,Takhistov:2017bpt}.

\section{Recommendations and Conclusions}
\label{sec:conclusions}

\subsection{Requirements for the Coming Decade}

\noindent
\textbf{Radio Timing of Pulsars.}
Until recently, most radio observatories observed only a small fraction of the known pulsar population due to limited and competitive resources. The recent operation of wide-field interferometers, such as the UTMOST and CHIME telescopes, are allowing for high-cadence observations of over one thousand pulsars on a near-daily basis. These observations allow for continued monitoring and improved measurements of relativistic deviations in pulsar-binary orbits, yielding the potential for additional mass and geometric measurements in the near future.

\begin{figure}[tb]
\centering
\includegraphics[width=0.94\textwidth]{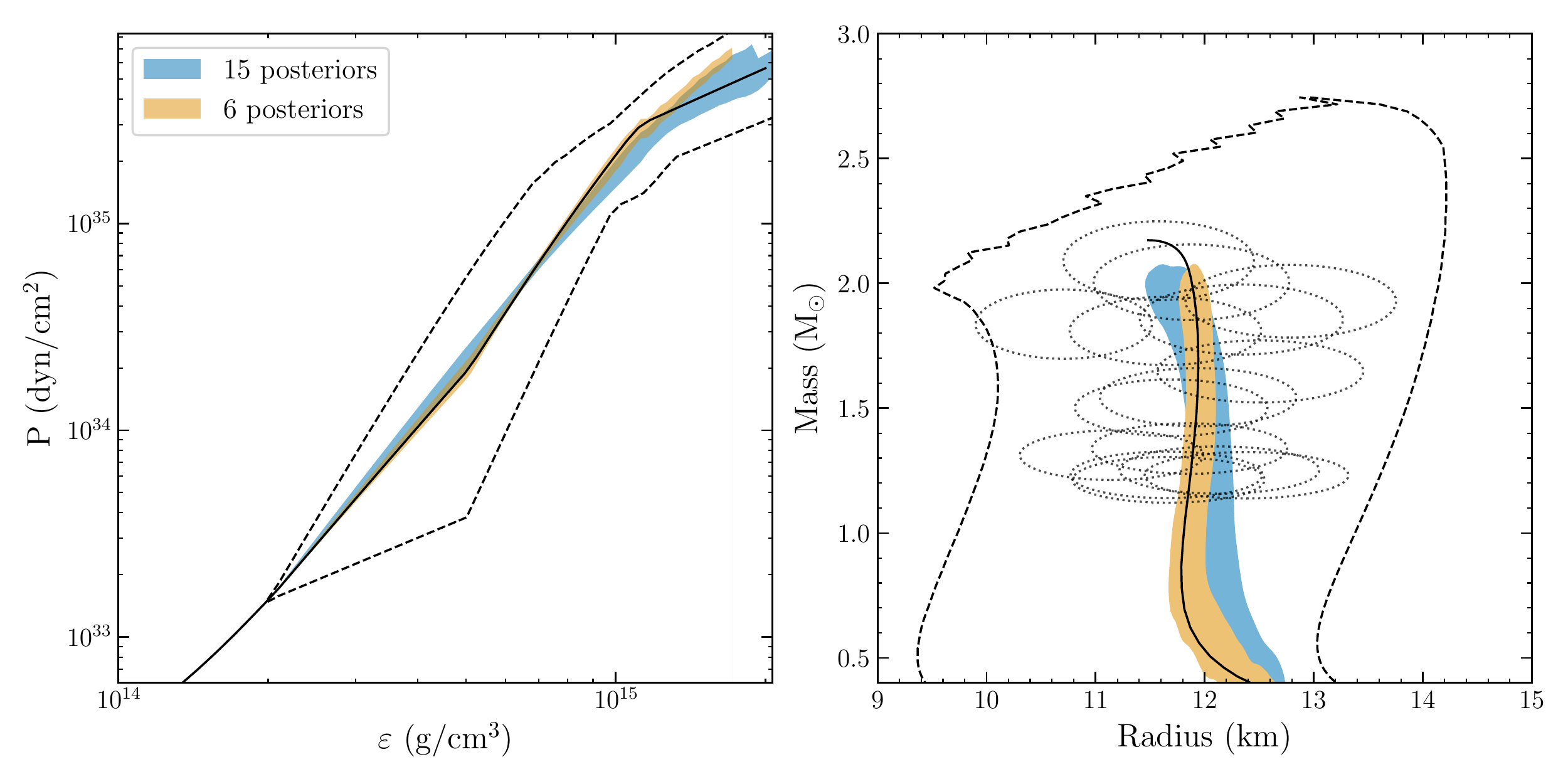}
\vspace{-0.2cm}
\caption{Simulated constraints on the EOS (pressure-density) and $M-R$ relation based on 15 expected measurements with \textit{STROBE-X}.  X-ray pulse profile modeling produces $M-R$ posteriors with $\pm 5$\% accuracy (dashed ellipses) scattered around an underlying $M-R$ relation (black line). 
The corresponding EOS is shown on the left hand panel, with the dashed lines indicating the range permitted by current models.  A piecewise-polytropic parameterization is assumed for the EOS with fixed transition densities.  Following the procedure outlined in \cite{Greif19}, the 1$\sigma$ constraints resulting from inference using these 15 measurements are shown by the blue band. The orange band shows the results for a choice of 6 of these stars, based on ensuring an even spread over the mass range $1.2 - 2.0$ M$_{\odot}$, for deeper observations that result in $\pm 2$\% accuracy posteriors (not shown). See \cite{2019arXiv190303035R} for further details.}	\label{fig:eosconstraints}
\end{figure}

\begin{table}[t!]
    \centering
    \small
    \begin{tabular}{l|c|c}
\hline
\hline
    \textbf{Technique}  & \textbf{NS source class} & \textbf{Observatories} \\
    \hline
    Pulse profile modeling  & rotation-powered MSPs & \textit{STROBE-X}, \textit{Athena} \\
                            & accretion-powered MSPs &        {\em STROBE-X}          \\
                            & bursting NS LMXBs  &        \textit{STROBE-X}     \\
                            \hline
    Extremely rapid rotation           &    accretion-powered MSPs    & \textit{STROBE-X}   \\
    \hline
    Radius expansion bursts  &    bursting NS LMXBs    & \textit{STROBE-X}   \\    
    \hline
    Absorption line spectroscopy     & bursting NS LMXBs & \textit{Athena}, \textit{Lynx} \\
    \hline
    Continuum spectroscopy     & quiescent NS LMXBs & \\
    Continuum spectroscopy     & young cooling NSs & \textit{AXIS}, \textit{Athena}, \textit{Lynx} \\
    \hline
    \end{tabular}
    \caption{\small{Summary of measurement techniques and NS source classes suitable for dense matter EOS constraints and the proposed X-ray observatories that have the capabilities to carry out these investigations.} }
    \label{tab:summary}
\end{table}

Forthcoming radio interferometers will provide greater sensitivity due to enlarged scales and frequency coverage. Recently, \cite{fcp+21} argued that timing of the high-mass PSR J0740+6620 with future observatories like the CHORD and DSA-2000 interferometers will allow for order-of-magnitude improvements on the pulsar mass precision, and likely allow for the measurement of apsidal motion in the system with several years of timing data. Once fully operational, the Square Kilometer Array will be able to detect the Shapiro time delay in at least 80\% of all known pulsar-binary systems with companion masses $\ge$ 0.1 M$_\odot$. The discovery space therefore remains open for exploration in the radio regime.

\textbf{X-ray Observatories in the 2020s and Beyond.}
Space-borne observations at X-ray energies offer various means for obtaining strong constraints on the allowed dense matter EOS, providing unique insight into the low temperature-high density region of the QCD phase diagram. While current telescopes have made important headway, they lack the required capabilities to fully exploit the information about the dense matter EOS encoded in the observed X-ray emission from NSs. This important undertaking requires a new generation of X-ray facilities with at least an order-of-magnitude improvement in sensitivity relative to current observatories, while also offering high time resolution required for effective studies of rapidly spinning NSs. 

The {\em Spectroscopic Time-Resolving Observatory for Broadband Energy X-rays} ({\em STROBE-X}) is a proposed probe-class X-ray (0.2--30 keV) timing and spectroscopy mission, with a 2--5\,m$^2$ collecting area, high spectral and temporal resolution, and rapid slew capabilities \cite{2019arXiv190303035R}. {\em STROBE-X} would enable several observational techniques to study the EOS \cite[see Table 1 and][for a review]{Watts16}.  The pulse profile modelling that can be done at present for a few pulsars with NICER (sufficient to provide a proof of concept, \cite{Bogdanov13}) would be possible for $\sim$20 pulsars with {\em STROBE-X} (Figure~\ref{fig:eosconstraints}), with more source classes becoming accessible through the hard X-ray band not presently covered.  

The {\em Advanced X-ray Imaging Satellite} (\textit{AXIS}) is a proposed probe-class soft X-ray (0.2--10 keV) mission, with $\sim$$1''$ angular resolution imaging capabilities and a collecting area of $\approx5,000$ cm$^{2}$ around 1 keV \cite{2018SPIE10699E..29M}, with possible deployment in the late 2020s or early 2030s. These design characteristics are optimal for spectroscopic studies of NS in crowded regions, such as sources in Galactic globular clusters and those embedded in supernova remnants and pulsar wind nebulae.

 The {\em Advanced Telescope for High-ENergy Astrophysics} ({\em Athena}) is the European Space Agency's next-generation flagship X-ray observatory, with key hardware contributions from NASA, slated for launch in the early 2030s  \cite{2013arXiv1306.2307N}. Its high sensitivity, high spectral resolution, and high count rate capability will enable studies of NSs where greatly improved spectroscopic sensitivity is desirable, such as for PRE burst systems and qLMXBs \cite{2013arXiv1306.2334M}.
 
 The \textit{Lynx X-ray Observatory} \cite{2018arXiv180909642T} is a concept soft X-ray facility for possible selection as a NASA Large Strategic Science Mission with a target launch date in the mid-2030s. It will have a novel combination of $\approx$2\,m$^2$ collecting area around 1 keV, high spectral resolution (through the use of a microcalorimeter and dispersion gratings), and grazing-incidence focusing optics that enable sub-arcsecond imaging. With these performance characteristics, {\em Lynx} would be able to produce potentially strong NS EOS constraints through high fidelity spectra of both quiescent and bursting NS LMXBs, especially those in the dense confines of globular clusters and the Galactic center where such systems are overabundant.

Table~\ref{tab:summary} summarizes the various techniques and NS source classes that can be employed to provide constraints on the dense matter EOS using X-ray observations, and the future planned observatories with the capabilities to accomplish the desired measurements. The ability to target multiple NS varieties is crucially important, as it enables verification of the measurement techniques, allowing characterization and mitigation of systematic errors.  For instance, a number of NSs exhibit both accretion-powered pulsations and thermonuclear burst oscillations, permitting pulse profile modelling for the same source using two different types of hot spot. Additionally, for the bursting sources, conducting spectroscopic modeling of the burst cooling tail can offer additional cross-checks of techniques \cite[see, e.g.][]{Nattila17}. By targeting more NSs, it will be possible to sample the EOS across a wider range of core densities.  This will map the EOS more fully, probing any potential phase transitions with finer resolution, and will move us out of the regime where EOS model parameter inference may be prior-dominated \citep[see for example][]{Greif19}.

{{\bf Double neutron star mergers.}} We call upon the community to build the multidisciplinary ingredients necessary for multi-messenger astronomy of compact binary systems.
This includes the installation of highly-sensitive observatories, such as 3G GW detectors and the next generation of EM telescopes, and the development of an efficient, reliable, and interdisciplinary hierarchical Bayesian framework for the interpretation of the growing number of upcoming multi-messenger sources.
Such a framework must account for selection effects within both GW and EM observations while considering all aspects of the source population simultaneously, including the fundamental aspects of strong interactions. 
By reliably extracting source properties (masses, radii, dynamical states, magnetic field structures, ejecta properties, etc.), this framework will enable reliable measurements of the dense matter EOS, the QCD phase diagram, and r-process nucleosynthesis. Ultimately, future observations of these astronomical systems at the extremes of gravity and density have the potential to reveal physics beyond the Standard Model. Investment in large scale computing is required to sufficiently cover the parameter space and accurate modelling of inspiral and post-merger waveform from BNS mergers.\\

In addition to the necessity for the aforementioned multi-messenger observational facilities, ample funding resources for further theoretical and terrestrial laboratory studies of matter in this regime are highly desirable. Finally, for all measurement techniques mentioned, the availability of state-of-the-art large-scale high performance computing resources is of essence. 

Collectively, electromagnetic and gravitational wave astrophysical measurements combined with terrestrial laboratory constraints, hold the promise to deliver definitive empirical constraints on the true nature of the densest matter in the Universe and the high-density, low-temperature region of the quantum chromodynamics phase space \cite{miller19,raaijmakers20,2020ApJ...892...55J,2020PhRvD.101l3007L}, potentially within a century of James Chadwick's discovery of the neutron. Furthermore, these efforts can, in principle, provide unique insight into other high-profile topics in modern physics such as the nature of dark matter, alternative theories of gravity, nucleon superfluidity and superconductivity, as well as a wide array of astrophysics, including r-process nucleosynthesis, primordial black holes, and stellar evolution. 

\clearpage

\bibliographystyle{aasjournal}
\addcontentsline{toc}{section}{References}
\bibliography{bogdanov_wp,fonseca_wp,ns-binaries-loi}

\begin{thebibliography}{}
\expandafter\ifx\csname natexlab\endcsname\relax\def\natexlab#1{#1}\fi
\providecommand{\url}[1]{\href{#1}{#1}}
\providecommand{\dodoi}[1]{doi:~\href{http://doi.org/#1}{\nolinkurl{#1}}}
\providecommand{\doeprint}[1]{\href{http://ascl.net/#1}{\nolinkurl{http://ascl.net/#1}}}
\providecommand{\doarXiv}[1]{\href{https://arxiv.org/abs/#1}{\nolinkurl{https://arxiv.org/abs/#1}}}

\bibitem[{Abbott {et~al.}(2017{\natexlab{a}})}]{Monitor:2017mdv}
Abbott, B., {et~al.} 2017{\natexlab{a}}, Astrophys. J. Lett., 848, L13,
  \dodoi{10.3847/2041-8213/aa920c}

\bibitem[{Abbott {et~al.}(2017{\natexlab{b}})}]{Abbott:2017xzu}
---. 2017{\natexlab{b}}, Nature, 551, 85, \dodoi{10.1038/nature24471}

\bibitem[{Abbott {et~al.}(2018{\natexlab{a}})}]{Aasi:2013wya}
---. 2018{\natexlab{a}}, Living Rev. Rel., 21, 3,
  \dodoi{10.1007/s41114-018-0012-9}

\bibitem[{Abbott {et~al.}(2020)}]{Aasi:2013wya-arxiv}
---. 2020, arXiv:1304.0670

\bibitem[{Abbott {et~al.}(2018{\natexlab{b}})Abbott, Abbott, Abbott, Acernese,
  Ackley, Adams, Adams, Addesso, Adhikari, Adya,
  {et~al.}}]{TheLIGOScientificCollaboration2018}
Abbott, B.~P., Abbott, R., Abbott, T.~D., {et~al.} 2018{\natexlab{b}}, Phys.
  Rev. Let., 121, 161101, \dodoi{10.1103/PhysRevLett.121.161101}

\bibitem[{{Abbott, B. P. et al.}(2017)}]{AbEA2017b}
{Abbott, B. P. et al.} 2017, Phys. Rev. Lett., 119, 161101,
  \dodoi{10.1103/PhysRevLett.119.161101}

\bibitem[{Ablyazimov {et~al.}(2017)}]{CBM:2016kpk}
Ablyazimov, T., {et~al.} 2017, Eur. Phys. J. A, 53, 60,
  \dodoi{10.1140/epja/i2017-12248-y}

\bibitem[{{Acernese} {et~al.}(2015){Acernese}, {Agathos}, {Agatsuma}, {Aisa},
  {Allemandou}, {Allocca}, {Amarni}, {Astone}, {Balestri}, {Ballardin}, \&
  et~al.}]{AcerneseAgathos2015}
{Acernese}, F., {Agathos}, M., {Agatsuma}, K., {et~al.} 2015, Classical and
  Quantum Gravity, 32, 024001, \dodoi{10.1088/0264-9381/32/2/024001}

\bibitem[{Ackley {et~al.}(2020)}]{Ackley:2020atn}
Ackley, K., {et~al.} 2020, arXiv, 2007.03128

\bibitem[{Adamczyk {et~al.}(2017)}]{STAR:2017sal}
Adamczyk, L., {et~al.} 2017, Phys. Rev. C, 96, 044904,
  \dodoi{10.1103/PhysRevC.96.044904}

\bibitem[{{Adhikari} {et~al.}(2021){Adhikari}, {Albataineh}, {Androic},
  {Aniol}, {Armstrong}, {Averett}, {Ayerbe Gayoso}, {Barcus}, {Bellini},
  {Beminiwattha}, {Benesch}, {Bhatt}, {Bhatta Pathak}, {Bhetuwal}, {Blaikie},
  {Campagna}, {Camsonne}, {Cates}, {Chen}, {Clarke}, {Cornejo}, {Covrig Dusa},
  {Datta}, {Deshpande}, {Dutta}, {Feldman}, {Fuchey}, {Gal}, {Gaskell},
  {Gautam}, {Gericke}, {Ghosh}, {Halilovic}, {Hansen}, {Hauenstein}, {Henry},
  {Horowitz}, {Jantzi}, {Jian}, {Johnston}, {Jones}, {Karki}, {Katugampola},
  {Keppel}, {King}, {King}, {Knauss}, {Kumar}, {Kutz}, {Lashley-Colthirst},
  {Leverick}, {Liu}, {Liyange}, {Malace}, {Mammei}, {Mammei}, {McCaughan},
  {McNulty}, {Meekins}, {Metts}, {Michaels}, {Mondal}, {Napolitano}, {Narayan},
  {Nikolaev}, {Rashad}, {Owen}, {Palatchi}, {Pan}, {Pandey}, {Park}, {Paschke},
  {Petrusky}, {Pitt}, {Premathilake}, {Puckett}, {Quinn}, {Radloff}, {Rahman},
  {Rathnayake}, {Reed}, {Reimer}, {Richards}, {Riordan}, {Roblin}, {Seeds},
  {Shahinyan}, {Souder}, {Tang}, {Thiel}, {Tian}, {Urciuoli}, {Wertz},
  {Wojtsekhowski}, {Yale}, {Ye}, {Yoon}, {Zec}, {Zhang}, {Zhang}, {Zheng}, \&
  {PREX Collaboration}}]{2021PhRvL.126q2502A}
{Adhikari}, D., {Albataineh}, H., {Androic}, D., {et~al.} 2021, \prl, 126,
  172502, \dodoi{10.1103/PhysRevLett.126.172502}

\bibitem[{{Agathos} {et~al.}(2020){Agathos}, {Zappa}, {Bernuzzi}, {Perego},
  {Breschi}, \& {Radice}}]{2020PhRvD.101d4006A}
{Agathos}, M., {Zappa}, F., {Bernuzzi}, S., {et~al.} 2020, Phys. Rev. D, 101,
  044006, \dodoi{10.1103/PhysRevD.101.044006}

\bibitem[{Aghamousa {et~al.}(2016)}]{Aghamousa:2016zmz}
Aghamousa, A., {et~al.} 2016, arXiv e-prints, 1611.00036.
\newblock \doarXiv{1611.00036}

\bibitem[{{Akmal} \& {Pandharipande}(1997)}]{Akmal97}
{Akmal}, A., \& {Pandharipande}, V.~R. 1997, \prc, 56, 2261,
  \dodoi{10.1103/PhysRevC.56.2261}

\bibitem[{Alex~Brown(2000)}]{PhysRevLett.85.5296}
Alex~Brown, B. 2000, Phys. Rev. Lett., 85, 5296,
  \dodoi{10.1103/PhysRevLett.85.5296}

\bibitem[{Alford {et~al.}(2019)Alford, Han, \&
  Schwenzer}]{alford_signatures_2019}
Alford, M.~G., Han, S., \& Schwenzer, K. 2019, J. Phys. G: Nucl. Part. Phys.,
  46, 114001, \dodoi{10.1088/1361-6471/ab337a}

\bibitem[{Alford {et~al.}(1998)Alford, Rajagopal, \& Wilczek}]{Alford:1997zt}
Alford, M.~G., Rajagopal, K., \& Wilczek, F. 1998, Phys. Lett. B, 422, 247,
  \dodoi{10.1016/S0370-2693(98)00051-3}

\bibitem[{Alford {et~al.}(2008)Alford, Schmitt, Rajagopal, \&
  Sch\"afer}]{Alford:2007xm}
Alford, M.~G., Schmitt, A., Rajagopal, K., \& Sch\"afer, T. 2008, Rev. Mod.
  Phys., 80, 1455, \dodoi{10.1103/RevModPhys.80.1455}

\bibitem[{{AlGendy} \& {Morsink}(2014)}]{Algendy14}
{AlGendy}, M., \& {Morsink}, S.~M. 2014, \apj, 791, 78,
  \dodoi{10.1088/0004-637X/791/2/78}

\bibitem[{{Allard} \& {Chamel}(2022)}]{allardchamel22}
{Allard}, V., \& {Chamel}, N. 2022, arXiv e-prints, arXiv:2203.08778.
\newblock \doarXiv{2203.08778}

\bibitem[{{Alpar} {et~al.}(1984){Alpar}, {Pines}, {Anderson}, \&
  {Shaham}}]{alparetal84}
{Alpar}, M.~A., {Pines}, D., {Anderson}, P.~W., \& {Shaham}, J. 1984, \apj,
  276, 325, \dodoi{10.1086/161616}

\bibitem[{{Anderson} \& {Itoh}(1975)}]{andersonitoh75}
{Anderson}, P.~W., \& {Itoh}, N. 1975, \nat, 256, 25, \dodoi{10.1038/256025a0}

\bibitem[{Annala {et~al.}(2020)Annala, Gorda, Kurkela, N{\"a}ttil{\"a}, \&
  Vuorinen}]{Annala:2019puf}
Annala, E., Gorda, T., Kurkela, A., N{\"a}ttil{\"a}, J., \& Vuorinen, A. 2020,
  Nat. Phys., 16, 907, \dodoi{10.1038/s41567-020-0914-9}

\bibitem[{Annala {et~al.}(2018)Annala, Gorda, Kurkela, \&
  Vuorinen}]{Annala:2017llu}
Annala, E., Gorda, T., Kurkela, A., \& Vuorinen, A. 2018, Phys. Rev. Lett.,
  120, 172703, \dodoi{10.1103/PhysRevLett.120.172703}

\bibitem[{Antoniadis {et~al.}(2013)}]{Antoniadis:2013pzd}
Antoniadis, J., {et~al.} 2013, Science, 340, 6131,
  \dodoi{10.1126/science.1233232}

\bibitem[{Aoki {et~al.}(2006)Aoki, Endrodi, Fodor, Katz, \&
  Szabo}]{Aoki:2006we}
Aoki, Y., Endrodi, G., Fodor, Z., Katz, S.~D., \& Szabo, K.~K. 2006, Nature,
  443, 675, \dodoi{10.1038/nature05120}

\bibitem[{Arcavi {et~al.}(2017)}]{Arcavi:2017xiz}
Arcavi, I., {et~al.} 2017, Nature, 551, 64, \dodoi{10.1038/nature24291}

\bibitem[{{Archibald} {et~al.}(2018){Archibald}, {Gusinskaia}, {Hessels},
  {Deller}, {Kaplan}, {Lorimer}, {Lynch}, {Ransom}, \& {Stairs}}]{agh+18}
{Archibald}, A.~M., {Gusinskaia}, N.~V., {Hessels}, J. W.~T., {et~al.} 2018,
  Nature, 559, 73, \dodoi{10.1038/s41586-018-0265-1}

\bibitem[{Arzoumanian {et~al.}(2018)}]{Arzoumanian:2017puf}
Arzoumanian, Z., {et~al.} 2018, Astrophys. J. Suppl., 235, 37,
  \dodoi{10.3847/1538-4365/aab5b0}

\bibitem[{{Aso} {et~al.}(2013){Aso}, {Michimura}, {Somiya}, {Ando}, {Miyakawa},
  {Sekiguchi}, {Tatsumi}, \& {Yamamoto}}]{AsoMichimura2013}
{Aso}, Y., {Michimura}, Y., {Somiya}, K., {et~al.} 2013, Phys.\,Rev.\,D, 88,
  043007, \dodoi{10.1103/PhysRevD.88.043007}

\bibitem[{{B. P. Abbott et al.}(2017)}]{AbEA2017f}
{B. P. Abbott et al.} 2017, The Astrophysical Journal Letters, 848, L12.
\newblock \url{http://stacks.iop.org/2041-8205/848/i=2/a=L12}

\bibitem[{{Baiotti}(2019)}]{2019PrPNP.10903714B}
{Baiotti}, L. 2019, Progress in Particle and Nuclear Physics, 109, 103714,
  \dodoi{10.1016/j.ppnp.2019.103714}

\bibitem[{{Barbieri} {et~al.}(2019){Barbieri}, {Salafia}, {Perego}, {Colpi}, \&
  {Ghirlanda}}]{BarbieriSalafia2019}
{Barbieri}, C., {Salafia}, O.~S., {Perego}, A., {Colpi}, M., \& {Ghirlanda}, G.
  2019, Astron. Astrophys., 625, A152, \dodoi{10.1051/0004-6361/201935443}

\bibitem[{{Baub{\"o}ck} {et~al.}(2013){Baub{\"o}ck}, {Berti}, {Psaltis}, \&
  {{\"O}zel}}]{Baubock13}
{Baub{\"o}ck}, M., {Berti}, E., {Psaltis}, D., \& {{\"O}zel}, F. 2013, \apj,
  777, 68, \dodoi{10.1088/0004-637X/777/1/68}

\bibitem[{{Bauswein} {et~al.}(2019){Bauswein}, {Bastian}, {Blaschke},
  {Chatziioannou}, {Clark}, {Fischer}, \& {Oertel}}]{2019PhRvL.122f1102B}
{Bauswein}, A., {Bastian}, N.-U.~F., {Blaschke}, D.~B., {et~al.} 2019, Phys.
  Rev. Let., 122, 061102, \dodoi{10.1103/PhysRevLett.122.061102}

\bibitem[{Bauswein {et~al.}(2013)Bauswein, Baumgarte, \&
  Janka}]{Bauswein:2013jpa}
Bauswein, A., Baumgarte, T.~W., \& Janka, H.~T. 2013, Phys. Rev. Lett., 111,
  131101, \dodoi{10.1103/PhysRevLett.111.131101}

\bibitem[{Bauswein {et~al.}(2021)Bauswein, Blacker, Lioutas, Soultanis,
  Vijayan, \& Stergioulas}]{Bauswein:2020xlt}
Bauswein, A., Blacker, S., Lioutas, G., {et~al.} 2021, Phys. Rev. D, 103,
  123004, \dodoi{10.1103/PhysRevD.103.123004}

\bibitem[{{Bauswein} \& {Janka}(2012)}]{Bauswein2012}
{Bauswein}, A., \& {Janka}, H.-T. 2012, Phys. Rev. Let., 108, 011101,
  \dodoi{10.1103/PhysRevLett.108.011101}

\bibitem[{{Bauswein} {et~al.}(2012){Bauswein}, {Janka}, {Hebeler}, \&
  {Schwenk}}]{Bauswein2012a}
{Bauswein}, A., {Janka}, H.-T., {Hebeler}, K., \& {Schwenk}, A. 2012, Phys.
  Rev. D, 86, 063001, \dodoi{10.1103/PhysRevD.86.063001}

\bibitem[{Bauswein {et~al.}(2017)Bauswein, Just, Janka, \&
  Stergioulas}]{Bauswein:2017vtn}
Bauswein, A., Just, O., Janka, H.-T., \& Stergioulas, N. 2017, Astrophys. J.
  Lett., 850, L34, \dodoi{10.3847/2041-8213/aa9994}

\bibitem[{{Bauswein} {et~al.}(2014){Bauswein}, {Stergioulas}, \&
  {Janka}}]{bsj14}
{Bauswein}, A., {Stergioulas}, N., \& {Janka}, H.-T. 2014, Phys. Rev. D, 90,
  023002, \dodoi{10.1103/PhysRevD.90.023002}

\bibitem[{{Bauswein} {et~al.}(2020){Bauswein}, {Blacker}, {Vijayan},
  {Stergioulas}, {Chatziioannou}, {Clark}, {Bastian}, {Blaschke}, {Cierniak},
  \& {Fischer}}]{2020arXiv200400846B}
{Bauswein}, A., {Blacker}, S., {Vijayan}, V., {et~al.} 2020, arXiv e-prints,
  arXiv:2004.00846.
\newblock \doarXiv{2004.00846}

\bibitem[{{Baym} {et~al.}(2018){Baym}, {Hatsuda}, {Kojo}, {Powell}, {Song}, \&
  {Takatsuka}}]{bhk+18}
{Baym}, G., {Hatsuda}, T., {Kojo}, T., {et~al.} 2018, Rep. Prog. Phys., 81,
  056902, \dodoi{10.1088/1361-6633/aaae14}

\bibitem[{{Baym} {et~al.}(1969){Baym}, {Pethick}, \& {Pines}}]{baymetal69}
{Baym}, G., {Pethick}, C., \& {Pines}, D. 1969, \nat, 224, 673,
  \dodoi{10.1038/224673a0}

\bibitem[{Baym {et~al.}(1971)Baym, Pethick, \& Sutherland}]{Baym:1971pw}
Baym, G., Pethick, C., \& Sutherland, P. 1971, Astrophys. J., 170, 299,
  \dodoi{10.1086/151216}

\bibitem[{{Bednarek} {et~al.}(2012){Bednarek}, {Haensel}, {Zdunik}, {Bejger},
  \& {Ma{\'n}ka}}]{Bednarek_etal_2012}
{Bednarek}, I., {Haensel}, P., {Zdunik}, J.~L., {Bejger}, M., \& {Ma{\'n}ka},
  R. 2012, \aap, 543, A157, \dodoi{10.1051/0004-6361/201118560}

\bibitem[{Bejger {et~al.}(2005)Bejger, Bulik, \&
  Haensel}]{bejger_constraints_2005}
Bejger, M., Bulik, T., \& Haensel, P. 2005, MNRAS, 364, 635,
  \dodoi{10.1111/j.1365-2966.2005.09575.x}

\bibitem[{Belczynski {et~al.}(2012)Belczynski, Wiktorowicz, Fryer, Holz, \&
  Kalogera}]{Belczynski_2012}
Belczynski, K., Wiktorowicz, G., Fryer, C.~L., Holz, D.~E., \& Kalogera, V.
  2012, The Astrophysical Journal, 757, 91, \dodoi{10.1088/0004-637x/757/1/91}

\bibitem[{{Bernuzzi}(2020)}]{2020arXiv200406419B}
{Bernuzzi}, S. 2020, arXiv e-prints, arXiv:2004.06419.
\newblock \doarXiv{2004.06419}

\bibitem[{Bernuzzi {et~al.}(2015)Bernuzzi, Dietrich, \&
  Nagar}]{Bernuzzi:2015rla}
Bernuzzi, S., Dietrich, T., \& Nagar, A. 2015, Phys. Rev. Lett., 115, 091101,
  \dodoi{10.1103/PhysRevLett.115.091101}

\bibitem[{{Bertone} \& {Fairbairn}(2008)}]{BertoneFairbairn2008}
{Bertone}, G., \& {Fairbairn}, M. 2008, Phys. Rev. D, 77, 043515,
  \dodoi{10.1103/PhysRevD.77.043515}

\bibitem[{Blacker {et~al.}(2020)Blacker, Bastian, Bauswein, Blaschke, Fischer,
  Oertel, Soultanis, \& Typel}]{Blacker:2020nlq}
Blacker, S., Bastian, N.-U.~F., Bauswein, A., {et~al.} 2020, Phys. Rev. D, 102,
  123023, \dodoi{10.1103/PhysRevD.102.123023}

\bibitem[{{Bogdanov}(2013)}]{Bogdanov13}
{Bogdanov}, S. 2013, \apj, 762, 96, \dodoi{10.1088/0004-637X/762/2/96}

\bibitem[{{Bogdanov} {et~al.}(2016){Bogdanov}, {Heinke}, {{\"O}zel}, \&
  {G{\"u}ver}}]{Bogdanov_et_al_2016}
{Bogdanov}, S., {Heinke}, C.~O., {{\"O}zel}, F., \& {G{\"u}ver}, T. 2016, \apj,
  831, 184, \dodoi{10.3847/0004-637X/831/2/184}

\bibitem[{{Bogdanov} {et~al.}(2019){Bogdanov}, {Lamb}, {Mahmoodifar}, {Miller},
  {Morsink}, {Riley}, {Strohmayer}, {Tung}, {Watts}, {Dittmann}, {Chakrabarty},
  {Guillot}, {Arzoumanian}, \& {Gendreau}}]{bogdanov19}
{Bogdanov}, S., {Lamb}, F.~K., {Mahmoodifar}, S., {et~al.} 2019, \apjl, 887,
  L26, \dodoi{10.3847/2041-8213/ab5968}

\bibitem[{Borhanian(2021)}]{Borhanian:2020ypi}
Borhanian, S. 2021, Class. Quant. Grav., 38, 175014,
  \dodoi{10.1088/1361-6382/ac1618}

\bibitem[{Bose {et~al.}(2018)Bose, Chakravarti, Rezzolla, Sathyaprakash, \&
  Takami}]{Bose2018}
Bose, S., Chakravarti, K., Rezzolla, L., Sathyaprakash, B.~S., \& Takami, K.
  2018, Phys. Rev. Let., 120, 031102, \dodoi{10.1103/PhysRevLett.120.031102}

\bibitem[{Bramante {et~al.}(2018)Bramante, Linden, \& Tsai}]{Bramante:2017ulk}
Bramante, J., Linden, T., \& Tsai, Y.-D. 2018, Phys. Rev. D, 97, 055016,
  \dodoi{10.1103/PhysRevD.97.055016}

\bibitem[{Breschi {et~al.}(2022{\natexlab{a}})Breschi, Bernuzzi, Chakravarti,
  Camilletti, Prakash, \& Perego}]{Breschi:2022xnc}
Breschi, M., Bernuzzi, S., Chakravarti, K., {et~al.} 2022{\natexlab{a}},
  Kilohertz Gravitational Waves From Binary Neutron Star Mergers:
  Numerical-relativity Informed Postmerger Model.
\newblock \doarXiv{2205.09112}

\bibitem[{Breschi {et~al.}(2019)Breschi, Bernuzzi, Zappa, Agathos, Perego,
  Radice, \& Nagar}]{Breschi:2019srl}
Breschi, M., Bernuzzi, S., Zappa, F., {et~al.} 2019, Phys. Rev. D, 100, 104029,
  \dodoi{10.1103/PhysRevD.100.104029}

\bibitem[{Breschi {et~al.}(2022{\natexlab{b}})Breschi, Gamba, Borhanian,
  Carullo, \& Bernuzzi}]{Breschi:2022ens}
Breschi, M., Gamba, R., Borhanian, S., Carullo, G., \& Bernuzzi, S.
  2022{\natexlab{b}}, {Kilohertz Gravitational Waves from Binary Neutron Star
  Mergers: Inference of Postmerger Signals with the Einstein Telescope}.
\newblock \doarXiv{2205.09979}

\bibitem[{{Burbidge}(1963)}]{Burbidge_1963}
{Burbidge}, G. 1963, \apj, 137, 995, \dodoi{10.1086/147575}

\bibitem[{{Capano} {et~al.}(2020){Capano}, {Tews}, {Brown}, {Margalit}, {De},
  {Kumar}, {Brown}, {Krishnan}, \& {Reddy}}]{2020NatAs.tmp...42C}
{Capano}, C.~D., {Tews}, I., {Brown}, S.~M., {et~al.} 2020, Nature Astronomy,
  \dodoi{10.1038/s41550-020-1014-6}

\bibitem[{Cardoso {et~al.}(2017)Cardoso, Franzin, Maselli, Pani, \&
  Raposo}]{Cardoso:2017cfl}
Cardoso, V., Franzin, E., Maselli, A., Pani, P., \& Raposo, G. 2017, Phys. Rev.
  D, 95, 084014, \dodoi{10.1103/PhysRevD.95.084014}

\bibitem[{{Carson} {et~al.}(2019){Carson}, {Chatziioannou}, {Haster}, {Yagi},
  \& {Yunes}}]{2019PhRvD..99h3016C}
{Carson}, Z., {Chatziioannou}, K., {Haster}, C.-J., {Yagi}, K., \& {Yunes}, N.
  2019, Phys. Rev. D, 99, 083016, \dodoi{10.1103/PhysRevD.99.083016}

\bibitem[{{Chamel}(2012)}]{chamel12}
{Chamel}, N. 2012, \prc, 85, 035801, \dodoi{10.1103/PhysRevC.85.035801}

\bibitem[{{Chang} {et~al.}(2005){Chang}, {Bildsten}, \&
  {Wasserman}}]{Chang_etal_2005}
{Chang}, P., {Bildsten}, L., \& {Wasserman}, I. 2005, \apj, 629, 998,
  \dodoi{10.1086/431730}

\bibitem[{{Chatziioannou}(2020)}]{2020arXiv200603168C}
{Chatziioannou}, K. 2020, arXiv e-prints, arXiv:2006.03168.
\newblock \doarXiv{2006.03168}

\bibitem[{Chatziioannou {et~al.}(2017)Chatziioannou, Clark, Bauswein,
  Millhouse, Littenberg, \& Cornish}]{Chatziioannou:2017ixj}
Chatziioannou, K., Clark, J.~A., Bauswein, A., {et~al.} 2017, Phys. Rev. D, 96,
  124035, \dodoi{10.1103/PhysRevD.96.124035}

\bibitem[{{Chatziioannou} \& {Han}(2020)}]{2020PhRvD.101d4019C}
{Chatziioannou}, K., \& {Han}, S. 2020, Phys. Rev. D, 101, 044019,
  \dodoi{10.1103/PhysRevD.101.044019}

\bibitem[{{Chatziioannou} {et~al.}(2015){Chatziioannou}, {Yagi}, {Klein},
  {Cornish}, \& {Yunes}}]{2015PhRvD..92j4008C}
{Chatziioannou}, K., {Yagi}, K., {Klein}, A., {Cornish}, N., \& {Yunes}, N.
  2015, Phys. Rev. D, 92, 104008, \dodoi{10.1103/PhysRevD.92.104008}

\bibitem[{{Chen} {et~al.}(2020){Chen}, {Johnson-McDaniel}, {Dietrich}, \&
  {Dudi}}]{2020PhRvD.101j3008C}
{Chen}, A., {Johnson-McDaniel}, N.~K., {Dietrich}, T., \& {Dudi}, R. 2020,
  Phys. Rev. D, 101, 103008, \dodoi{10.1103/PhysRevD.101.103008}

\bibitem[{Cherman {et~al.}(2019)Cherman, Sen, \& Yaffe}]{Cherman:2018jir}
Cherman, A., Sen, S., \& Yaffe, L.~G. 2019, Phys. Rev. D, 100, 034015,
  \dodoi{10.1103/PhysRevD.100.034015}

\bibitem[{{Ciarcelluti} \& {Sandin}(2011)}]{CiarcellutiSandin2011}
{Ciarcelluti}, P., \& {Sandin}, F. 2011, Physics Letters B, 695, 19,
  \dodoi{10.1016/j.physletb.2010.11.021}

\bibitem[{{Clark} {et~al.}(2014){Clark}, {Bauswein}, {Cadonati}, {Janka},
  {Pankow}, \& {Stergioulas}}]{Clark2014}
{Clark}, J., {Bauswein}, A., {Cadonati}, L., {et~al.} 2014, Phys. Rev. D, 90,
  062004, \dodoi{10.1103/PhysRevD.90.062004}

\bibitem[{{Clark} {et~al.}(2016){Clark}, {Bauswein}, {Stergioulas}, \&
  {Shoemaker}}]{Clark2016}
{Clark}, J.~A., {Bauswein}, A., {Stergioulas}, N., \& {Shoemaker}, D. 2016,
  Classical and Quantum Gravity, 33, 085003,
  \dodoi{10.1088/0264-9381/33/8/085003}

\bibitem[{Clough {et~al.}(2018)Clough, Dietrich, \& Niemeyer}]{Clough:2018exo}
Clough, K., Dietrich, T., \& Niemeyer, J.~C. 2018, Phys. Rev. D, 98, 083020,
  \dodoi{10.1103/PhysRevD.98.083020}

\bibitem[{Coughlin {et~al.}(2019)Coughlin, Dietrich, Margalit, \&
  Metzger}]{CoDi2018b}
Coughlin, M.~W., Dietrich, T., Margalit, B., \& Metzger, B.~D. 2019, MNRAS:
  Letters, 489, L91, \dodoi{10.1093/mnrasl/slz133}

\bibitem[{Coughlin {et~al.}(2020{\natexlab{a}})Coughlin, Dietrich, Heinzel,
  Khetan, Antier, Bulla, Christensen, Coulter, \& Foley}]{Coughlin:2019vtv}
Coughlin, M.~W., Dietrich, T., Heinzel, J., {et~al.} 2020{\natexlab{a}}, Phys.
  Rev. Res., 2, 022006, \dodoi{10.1103/PhysRevResearch.2.022006}

\bibitem[{Coughlin {et~al.}(2020{\natexlab{b}})Coughlin, Antier, Dietrich,
  Foley, Heinzel, Bulla, Christensen, Coulter, Issa, \&
  Khetan}]{Coughlin:2020ozl}
Coughlin, M.~W., Antier, S., Dietrich, T., {et~al.} 2020{\natexlab{b}}, Nature
  Commun., 11, 4129, \dodoi{10.1038/s41467-020-17998-5}

\bibitem[{Coughlin {et~al.}(2020{\natexlab{c}})}]{Coughlin:2020fwx}
Coughlin, M.~W., {et~al.} 2020{\natexlab{c}}, Mon. Not. Roy. Astron. Soc., 497,
  1181, \dodoi{10.1093/mnras/staa1925}

\bibitem[{Coughlin {et~al.}(2020{\natexlab{d}})Coughlin, Dietrich, Antier,
  Bulla, Foucart, Hotokezaka, Raaijmakers, Hinderer, \&
  Nissanke}]{Coughlin:2019zqi}
Coughlin, M.~W., Dietrich, T., Antier, S., {et~al.} 2020{\natexlab{d}}, Mon.
  Not. Roy. Astron. Soc., 492, 863, \dodoi{10.1093/mnras/stz3457}

\bibitem[{Coulter {et~al.}(2017)}]{Coulter:2017wya}
Coulter, D., {et~al.} 2017, Science, 358, 1556, \dodoi{10.1126/science.aap9811}

\bibitem[{{Cromartie} {et~al.}(2020){Cromartie}, {Fonseca}, {Ransom},
  {Demorest}, {Arzoumanian}, {Blumer}, {Brook}, {DeCesar}, {Dolch}, {Ellis},
  {Ferdman}, {Ferrara}, {Garver-Daniels}, {Gentile}, {Jones}, {Lam}, {Lorimer},
  {Lynch}, {McLaughlin}, {Ng}, {Nice}, {Pennucci}, {Spiewak}, {Stairs},
  {Stovall}, {Swiggum}, \& {Zhu}}]{cfr+20}
{Cromartie}, H.~T., {Fonseca}, E., {Ransom}, S.~M., {et~al.} 2020, Nature
  Astronomy, 4, 72, \dodoi{10.1038/s41550-019-0880-2}

\bibitem[{{Damen} {et~al.}(1990){Damen}, {Magnier}, {Lewin}, {Tan}, {Penninx},
  \& {van Paradijs}}]{Damen_etal_1990}
{Damen}, E., {Magnier}, E., {Lewin}, W.~H.~G., {et~al.} 1990, \aap, 237, 103

\bibitem[{{Damour} \& {Schafer}(1988)}]{ds88}
{Damour}, T., \& {Schafer}, G. 1988, Nuovo Cimento B Serie, 101B, 127,
  \dodoi{10.1007/BF02828697}

\bibitem[{De {et~al.}(2018)De, Finstad, Lattimer, Brown, Berger, \&
  Biwer}]{De2018}
De, S., Finstad, D., Lattimer, J.~M., {et~al.} 2018, Phys. Rev. Let., 121,
  091102, \dodoi{10.1103/PhysRevLett.121.091102}

\bibitem[{{de Lavallaz} \& {Fairbairn}(2010)}]{deLavallazFairbairn2010}
{de Lavallaz}, A., \& {Fairbairn}, M. 2010, Phys. Rev. D, 81, 123521,
  \dodoi{10.1103/PhysRevD.81.123521}

\bibitem[{{De Pietri} {et~al.}(2019){De Pietri}, {Drago}, {Feo}, {Pagliara},
  {Pasquali}, {Traversi}, \& {Wiktorowicz}}]{2019ApJ...881..122D}
{De Pietri}, R., {Drago}, A., {Feo}, A., {et~al.} 2019, Astrophys. J., 881,
  122, \dodoi{10.3847/1538-4357/ab2fd0}

\bibitem[{{Del Pozzo} {et~al.}(2013){Del Pozzo}, {Li}, {Agathos}, {Van Den
  Broeck}, \& {Vitale}}]{2013PhRvL.111g1101D}
{Del Pozzo}, W., {Li}, T. G.~F., {Agathos}, M., {Van Den Broeck}, C., \&
  {Vitale}, S. 2013, Phys. Rev. Let., 111, 071101,
  \dodoi{10.1103/PhysRevLett.111.071101}

\bibitem[{Del~Pozzo {et~al.}(2017)Del~Pozzo, Li, \&
  Messenger}]{PhysRevD.95.043502}
Del~Pozzo, W., Li, T. G.~F., \& Messenger, C. 2017, Phys. Rev. D, 95, 043502,
  \dodoi{10.1103/PhysRevD.95.043502}

\bibitem[{{Deliyergiyev} {et~al.}(2019){Deliyergiyev}, {Del Popolo}, {Tolos},
  {Le Delliou}, {Lee}, \& {Burgio}}]{DeliyergiyevDelPopolo2019}
{Deliyergiyev}, M., {Del Popolo}, A., {Tolos}, L., {et~al.} 2019, Phys. Rev. D,
  99, 063015, \dodoi{10.1103/PhysRevD.99.063015}

\bibitem[{{Demorest} {et~al.}(2010){Demorest}, {Pennucci}, {Ransom}, {Roberts},
  \& {Hessels}}]{dem10}
{Demorest}, P.~B., {Pennucci}, T., {Ransom}, S.~M., {Roberts}, M.~S.~E., \&
  {Hessels}, J.~W.~T. 2010, Nature, 467, 1081, \dodoi{10.1038/nature09466}

\bibitem[{Dietrich \& Clough(2019)}]{Dietrich:2019shr}
Dietrich, T., \& Clough, K. 2019, Phys. Rev. D, 100, 083005,
  \dodoi{10.1103/PhysRevD.100.083005}

\bibitem[{Dietrich {et~al.}(2020)Dietrich, Coughlin, Pang, Bulla, Heinzel,
  Issa, Tews, \& Antier}]{Dietrich:2020lps}
Dietrich, T., Coughlin, M.~W., Pang, P. T.~H., {et~al.} 2020, arXiv e-prints,
  arxiv:2002.11355.
\newblock \doarXiv{2002.11355}

\bibitem[{Dietrich {et~al.}(2019)Dietrich, Day, Clough, Coughlin, \&
  Niemeyer}]{Dietrich:2018jov}
Dietrich, T., Day, F., Clough, K., Coughlin, M., \& Niemeyer, J. 2019, Mon.
  Not. Roy. Astron. Soc., 483, 908, \dodoi{10.1093/mnras/sty3158}

\bibitem[{{Dietrich} {et~al.}(2020){Dietrich}, {Hinderer}, \&
  {Samajdar}}]{2020arXiv200402527D}
{Dietrich}, T., {Hinderer}, T., \& {Samajdar}, A. 2020, arXiv e-prints,
  arXiv:2004.02527.
\newblock \doarXiv{2004.02527}

\bibitem[{Drischler {et~al.}(2021)Drischler, Holt, \&
  Wellenhofer}]{doi:10.1146/annurev-nucl-102419-041903}
Drischler, C., Holt, J., \& Wellenhofer, C. 2021, Annual Review of Nuclear and
  Particle Science, 71, 403, \dodoi{10.1146/annurev-nucl-102419-041903}

\bibitem[{Easter {et~al.}(2020)Easter, Ghonge, Lasky, Casey, Clark,
  Hernandez~Vivanco, \& Chatziioannou}]{Easter:2020ifj}
Easter, P.~J., Ghonge, S., Lasky, P.~D., {et~al.} 2020, Phys. Rev. D, 102,
  043011, \dodoi{10.1103/PhysRevD.102.043011}

\bibitem[{{Essick} \& {Landry}(2020)}]{2020arXiv200701372E}
{Essick}, R., \& {Landry}, P. 2020, arXiv e-prints, arXiv:2007.01372.
\newblock \doarXiv{2007.01372}

\bibitem[{Essick {et~al.}(2020)Essick, Landry, \& Holz}]{Essick:2019ldf}
Essick, R., Landry, P., \& Holz, D.~E. 2020, Phys. Rev. D, 101, 063007,
  \dodoi{10.1103/PhysRevD.101.063007}

\bibitem[{{Essick} {et~al.}(2020){Essick}, {Tews}, {Landry}, {Reddy}, \&
  {Holz}}]{2020arXiv200407744E}
{Essick}, R., {Tews}, I., {Landry}, P., {Reddy}, S., \& {Holz}, D.~E. 2020,
  arXiv e-prints, arXiv:2004.07744.
\newblock \doarXiv{2004.07744}

\bibitem[{Essick {et~al.}(2021)Essick, Tews, Landry, \&
  Schwenk}]{Essick:2021kjb}
Essick, R., Tews, I., Landry, P., \& Schwenk, A. 2021, Phys. Rev. Lett., 127,
  192701, \dodoi{10.1103/PhysRevLett.127.192701}

\bibitem[{Essick {et~al.}(2015)Essick, Vitale, Katsavounidis, Vedovato, \&
  Klimenko}]{Essick:2014wwa}
Essick, R., Vitale, S., Katsavounidis, E., Vedovato, G., \& Klimenko, S. 2015,
  Astrophys. J., 800, 81, \dodoi{10.1088/0004-637X/800/2/81}

\bibitem[{Fairhurst(2009)}]{Fairhurst:2009tc}
Fairhurst, S. 2009, New J. Phys., 11, 123006,
  \dodoi{10.1088/1367-2630/11/12/123006}

\bibitem[{Farmer {et~al.}(2020)Farmer, Renzo, de~Mink, Fishbach, \&
  Justham}]{Farmer:2020xne}
Farmer, R., Renzo, M., de~Mink, S., Fishbach, M., \& Justham, S. 2020, arXiv
  e-prints, arXiv:2006.06678.
\newblock \doarXiv{2006.06678}

\bibitem[{Farr {et~al.}(2019)Farr, Fishbach, Ye, \& Holz}]{Farr_2019}
Farr, W.~M., Fishbach, M., Ye, J., \& Holz, D.~E. 2019, The Astrophysical
  Journal, 883, L42, \dodoi{10.3847/2041-8213/ab4284}

\bibitem[{{Fasano} {et~al.}(2020){Fasano}, {Wong}, {Maselli}, {Berti},
  {Ferrari}, \& {Sathyaprakash}}]{2020PhRvD.102b3025F}
{Fasano}, M., {Wong}, K. W.~K., {Maselli}, A., {et~al.} 2020, Phys. Rev. D,
  102, 023025, \dodoi{10.1103/PhysRevD.102.023025}

\bibitem[{Fattoyev \& Piekarewicz(2010)}]{fattoyev_sensitivity_2010}
Fattoyev, F.~J., \& Piekarewicz, J. 2010, Phys. Rev. C, 82, 025810,
  \dodoi{10.1103/PhysRevC.82.025810}

\bibitem[{Fattoyev {et~al.}(2018)Fattoyev, Piekarewicz, \&
  Horowitz}]{Fattoyev2018}
Fattoyev, F.~J., Piekarewicz, J., \& Horowitz, C.~J. 2018, Phys. Rev. Let.,
  120, 172702, \dodoi{10.1103/PhysRevLett.120.172702}

\bibitem[{Fodor \& Katz(2004)}]{Fodor:2004nz}
Fodor, Z., \& Katz, S.~D. 2004, JHEP, 04, 050,
  \dodoi{10.1088/1126-6708/2004/04/050}

\bibitem[{{Fonseca} {et~al.}(2014){Fonseca}, {Stairs}, \& {Thorsett}}]{fst10}
{Fonseca}, E., {Stairs}, I.~H., \& {Thorsett}, S.~E. 2014, ApJ, 787, 82,
  \dodoi{10.1088/0004-637X/787/1/82}

\bibitem[{{Fonseca} {et~al.}(2021){Fonseca}, {Cromartie}, {Pennucci}, {Ray},
  {Kirichenko}, {Ransom}, {Demorest}, {Stairs}, {Arzoumanian}, {Guillemot},
  {Parthasarathy}, {Kerr}, {Cognard}, {Baker}, {Blumer}, {Brook}, {DeCesar},
  {Dolch}, {Dong}, {Ferrara}, {Fiore}, {Garver-Daniels}, {Good}, {Jennings},
  {Jones}, {Kaspi}, {Lam}, {Lorimer}, {Luo}, {McEwen}, {McKee}, {McLaughlin},
  {McMann}, {Meyers}, {Naidu}, {Ng}, {Nice}, {Pol}, {Radovan},
  {Shapiro-Albert}, {Tan}, {Tendulkar}, {Swiggum}, {Wahl}, \& {Zhu}}]{fcp+21}
{Fonseca}, E., {Cromartie}, H.~T., {Pennucci}, T.~T., {et~al.} 2021, ApJL, 915,
  L12, \dodoi{10.3847/2041-8213/ac03b8}

\bibitem[{Fryer \& Kalogera(2001)}]{Fryer_2001}
Fryer, C.~L., \& Kalogera, V. 2001, The Astrophysical Journal, 554, 548,
  \dodoi{10.1086/321359}

\bibitem[{Fujimoto {et~al.}(2022)Fujimoto, Fukushima, McLerran, \&
  Praszalowicz}]{Fujimoto:2022ohj}
Fujimoto, Y., Fukushima, K., McLerran, L.~D., \& Praszalowicz, M. 2022.
\newblock \doarXiv{2207.06753}

\bibitem[{Fukushima \& Hatsuda(2011)}]{Fukushima:2010bq}
Fukushima, K., \& Hatsuda, T. 2011, Rept. Prog. Phys., 74, 014001,
  \dodoi{10.1088/0034-4885/74/1/014001}

\bibitem[{Fuller {et~al.}(2017)Fuller, Kusenko, \& Takhistov}]{Fuller:2017uyd}
Fuller, G.~M., Kusenko, A., \& Takhistov, V. 2017, Phys. Rev. Lett., 119,
  061101, \dodoi{10.1103/PhysRevLett.119.061101}

\bibitem[{{Gendreau} {et~al.}(2016){Gendreau}, {Arzoumanian}, {Adkins},
  {Albert}, {Anders}, {Aylward}, {Baker}, {Balsamo}, {Bamford}, {Benegalrao},
  {Berry}, {Bhalwani}, {Black}, {Blaurock}, {Bronke}, {Brown}, {Budinoff},
  {Cantwell}, {Cazeau}, {Chen}, {Clement}, {Colangelo}, {Coleman},
  {Coopersmith}, {Dehaven}, {Doty}, {Egan}, {Enoto}, {Fan}, {Ferro}, {Foster},
  {Galassi}, {Gallo}, {Green}, {Grosh}, {Ha}, {Hasouneh}, {Heefner}, {Hestnes},
  {Hoge}, {Jacobs}, {J{\o}rgensen}, {Kaiser}, {Kellogg}, {Kenyon}, {Koenecke},
  {Kozon}, {LaMarr}, {Lambertson}, {Larson}, {Lentine}, {Lewis}, {Lilly},
  {Liu}, {Malonis}, {Manthripragada}, {Markwardt}, {Matonak}, {Mcginnis},
  {Miller}, {Mitchell}, {Mitchell}, {Mohammed}, {Monroe}, {Montt de Garcia},
  {Mul{\'e}}, {Nagao}, {Ngo}, {Norris}, {Norwood}, {Novotka}, {Okajima},
  {Olsen}, {Onyeachu}, {Orosco}, {Peterson}, {Pevear}, {Pham}, {Pollard},
  {Pope}, {Powers}, {Powers}, {Price}, {Prigozhin}, {Ramirez}, {Reid},
  {Remillard}, {Rogstad}, {Rosecrans}, {Rowe}, {Sager}, {Sanders}, {Savadkin},
  {Saylor}, {Schaeffer}, {Schweiss}, {Semper}, {Serlemitsos}, {Shackelford},
  {Soong}, {Struebel}, {Vezie}, {Villasenor}, {Winternitz}, {Wofford},
  {Wright}, {Yang}, \& {Yu}}]{2016SPIE.9905E..1HG}
{Gendreau}, K.~C., {Arzoumanian}, Z., {Adkins}, P.~W., {et~al.} 2016, in
  \procspie, Vol. 9905, Space Telescopes and Instrumentation 2016: Ultraviolet
  to Gamma Ray, 99051H, \dodoi{10.1117/12.2231304}

\bibitem[{{Glampedakis} \& {Gualtieri}(2018)}]{2018ASSL..457..673G}
{Glampedakis}, K., \& {Gualtieri}, L. 2018, Astrophysics and Space Science
  Library, Vol. 457, {Gravitational Waves from Single Neutron Stars: An
  Advanced Detector Era Survey} (Springer Nature), 673,
  \dodoi{10.1007/978-3-319-97616-7_12}

\bibitem[{Goldstein {et~al.}(2017)}]{Goldstein:2017mmi}
Goldstein, A., {et~al.} 2017, Astrophys. J. Lett., 848, L14,
  \dodoi{10.3847/2041-8213/aa8f41}

\bibitem[{{Gonzalez} {et~al.}(2011){Gonzalez}, {Stairs}, {Ferdman}, {Freire},
  {Nice}, {Demorest}, {Ransom}, {Kramer}, {Camilo}, {Hobbs}, {Manchester}, \&
  {Lyne}}]{gsf+11}
{Gonzalez}, M.~E., {Stairs}, I.~H., {Ferdman}, R.~D., {et~al.} 2011, ApJ, 743,
  102, \dodoi{10.1088/0004-637X/743/2/102}

\bibitem[{Gorda {et~al.}(2021)Gorda, Kurkela, Paatelainen, S\"appi, \&
  Vuorinen}]{Gorda:2021znl}
Gorda, T., Kurkela, A., Paatelainen, R., S\"appi, S., \& Vuorinen, A. 2021,
  Phys. Rev. Lett., 127, 162003, \dodoi{10.1103/PhysRevLett.127.162003}

\bibitem[{Gorda \& S\"appi(2022)}]{Gorda:2021gha}
Gorda, T., \& S\"appi, S. 2022, Phys. Rev. D, 105, 114005,
  \dodoi{10.1103/PhysRevD.105.114005}

\bibitem[{Greif {et~al.}(2020)Greif, Hebeler, Lattimer, Pethick, \&
  Schwenk}]{greif_equation_2020}
Greif, S.~K., Hebeler, K., Lattimer, J.~M., Pethick, C.~J., \& Schwenk, A.
  2020, arXiv:2005.14164.
\newblock \url{http://arxiv.org/abs/2005.14164}

\bibitem[{{Greif} {et~al.}(2019){Greif}, {Raaijmakers}, {Hebeler}, {Schwenk},
  \& {Watts}}]{Greif19}
{Greif}, S.~K., {Raaijmakers}, G., {Hebeler}, K., {Schwenk}, A., \& {Watts},
  A.~L. 2019, \mnras, 485, 5363, \dodoi{10.1093/mnras/stz654}

\bibitem[{{Gresham} \& {Zurek}(2019)}]{GreshamZurek2019}
{Gresham}, M.~I., \& {Zurek}, K.~M. 2019, Phys. Rev. D, 99, 083008,
  \dodoi{10.1103/PhysRevD.99.083008}

\bibitem[{{Guerra Chaves} \& {Hinderer}(2019)}]{2019JPhG...46l3002G}
{Guerra Chaves}, A., \& {Hinderer}, T. 2019, Journal of Physics G Nuclear
  Physics, 46, 123002, \dodoi{10.1088/1361-6471/ab45be}

\bibitem[{{Guillot} {et~al.}(2013){Guillot}, {Servillat}, {Webb}, \&
  {Rutledge}}]{Guillot_et_al_2013}
{Guillot}, S., {Servillat}, M., {Webb}, N.~A., \& {Rutledge}, R.~E. 2013, \apj,
  772, 7, \dodoi{10.1088/0004-637X/772/1/7}

\bibitem[{Gupta {et~al.}(2019)Gupta, Fox, Sathyaprakash, \&
  Schutz}]{Gupta:2019okl}
Gupta, A., Fox, D., Sathyaprakash, B.~S., \& Schutz, B.~F. 2019, The
  Astrophysical Journal, 886, 71, \dodoi{10.3847/1538-4357/ab4c92}

\bibitem[{{Gusakov} {et~al.}(2004){Gusakov}, {Kaminker}, {Yakovlev}, \&
  {Gnedin}}]{gusakovetal04}
{Gusakov}, M.~E., {Kaminker}, A.~D., {Yakovlev}, D.~G., \& {Gnedin}, O.~Y.
  2004, \aap, 423, 1063, \dodoi{10.1051/0004-6361:20041006}

\bibitem[{{G{\"u}ver} {et~al.}(2012){G{\"u}ver}, {{\"O}zel}, \&
  {Psaltis}}]{Guver_etal_2012}
{G{\"u}ver}, T., {{\"O}zel}, F., \& {Psaltis}, D. 2012, \apj, 747, 77,
  \dodoi{10.1088/0004-637X/747/1/77}

\bibitem[{{Haensel} {et~al.}(2009){Haensel}, {Zdunik}, {Bejger}, \&
  {Lattimer}}]{Haensel09}
{Haensel}, P., {Zdunik}, J.~L., {Bejger}, M., \& {Lattimer}, J.~M. 2009, \aap,
  502, 605, \dodoi{10.1051/0004-6361/200811605}

\bibitem[{{Hanauske} {et~al.}(2019){Hanauske}, {Bovard}, {Most}, {Papenfort},
  {Steinheimer}, {Motornenko}, {Vovchenko}, {Dexheimer}, {Schramm}, \&
  {St{\"o}cker}}]{2019Univ....5..156H}
{Hanauske}, M., {Bovard}, L., {Most}, E., {et~al.} 2019, Universe, 5, 156,
  \dodoi{10.3390/universe5060156}

\bibitem[{{Hebeler} {et~al.}(2013){Hebeler}, {Lattimer}, {Pethick}, \&
  {Schwenk}}]{hebeler13}
{Hebeler}, K., {Lattimer}, J.~M., {Pethick}, C.~J., \& {Schwenk}, A. 2013,
  \apj, 773, 11, \dodoi{10.1088/0004-637X/773/1/11}

\bibitem[{Hebeler \& Schwenk(2010)}]{Hebeler:2009iv}
Hebeler, K., \& Schwenk, A. 2010, Phys. Rev. C, 82, 014314,
  \dodoi{10.1103/PhysRevC.82.014314}

\bibitem[{{Heinke}(2013)}]{heinke13}
{Heinke}, C.~O. 2013, Journal of Physics Conference Series, 432, 012001,
  \dodoi{10.1088/1742-6596/432/1/012001}

\bibitem[{{Hernandez Vivanco} {et~al.}(2019){Hernandez Vivanco}, {Smith},
  {Thrane}, {Lasky}, {Talbot}, \& {Raymond}}]{2019PhRvD.100j3009H}
{Hernandez Vivanco}, F., {Smith}, R., {Thrane}, E., {et~al.} 2019, Phys. Rev.
  D, 100, 103009, \dodoi{10.1103/PhysRevD.100.103009}

\bibitem[{Hild {et~al.}(2011)}]{Hild:2010id}
Hild, S., {et~al.} 2011, Class. Quant. Grav., 28, 094013,
  \dodoi{10.1088/0264-9381/28/9/094013}

\bibitem[{{Hinderer} {et~al.}(2018){Hinderer}, {Rezzolla}, \&
  {Baiotti}}]{2018ASSL..457..575H}
{Hinderer}, T., {Rezzolla}, L., \& {Baiotti}, L. 2018, Astrophysics and Space
  Science Library, Vol. 457, {Gravitational Waves from Merging Binary
  Neutron-Star Systems} (Springer Nature), 575,
  \dodoi{10.1007/978-3-319-97616-7_10}

\bibitem[{Hinderer {et~al.}(2019)}]{Hinderer:2018pei}
Hinderer, T., {et~al.} 2019, Phys. Rev. D, 100, 06321,
  \dodoi{10.1103/PhysRevD.100.063021}

\bibitem[{{Ho}(2018)}]{2018RSPTA.37670285H}
{Ho}, W. C.~G. 2018, Philosophical Transactions of the Royal Society of London
  Series A, 376, 20170285, \dodoi{10.1098/rsta.2017.0285}

\bibitem[{{Ho} \& {Andersson}(2012)}]{hoandersson12}
{Ho}, W. C.~G., \& {Andersson}, N. 2012, Nature Physics, 8, 787,
  \dodoi{10.1038/nphys2424}

\bibitem[{Hotokezaka {et~al.}(2011)Hotokezaka, Kyutoku, Okawa, Shibata, \&
  Kiuchi}]{Hotokezaka:2011dh}
Hotokezaka, K., Kyutoku, K., Okawa, H., Shibata, M., \& Kiuchi, K. 2011, Phys.
  Rev. D, 83, 124008, \dodoi{10.1103/PhysRevD.83.124008}

\bibitem[{{Hu} {et~al.}(2020){Hu}, {Kramer}, {Wex}, {Champion}, \&
  {Kehl}}]{hkw+20}
{Hu}, H., {Kramer}, M., {Wex}, N., {Champion}, D.~J., \& {Kehl}, M.~S. 2020,
  MNRAS, 497, 3118, \dodoi{10.1093/mnras/staa2107}

\bibitem[{Huth {et~al.}(2022)}]{Huth:2021bsp}
Huth, S., {et~al.} 2022, Nature, 606, 276, \dodoi{10.1038/s41586-022-04750-w}

\bibitem[{Huxford {et~al.}(2022)Huxford, Kashyap, Borhania, Dhani, \&
  Sathyaprakash}]{Huxford:2022ab}
Huxford, R., Kashyap, R., Borhania, S., Dhani, A., \& Sathyaprakash, B. 2022,
  Dense matter equation of state with Cosmic Explorer and Einstein Telescope
  \emph{in preparation}.

\bibitem[{Ivezi\'c {et~al.}(2019)}]{Ivezic:2008fe}
Ivezi\'c, v., {et~al.} 2019, Astrophys. J., 873, 111,
  \dodoi{10.3847/1538-4357/ab042c}

\bibitem[{{Iyer} {et~al.}(2011){Iyer}, {Souradeep}, Unnikrishnan, Dhurandhar,
  Raja, Kumar, \& Sengupta}]{IYER-LIGOINDIA}
{Iyer}, B., {Souradeep}, T., Unnikrishnan, C., {et~al.} 2011, LIGO-India Tech.
  rep., LIGO-M1100296.
\newblock \url{https://dcc.ligo.org/LIGO-M1100296/public}

\bibitem[{{Jiang} {et~al.}(2020){Jiang}, {Tang}, {Wang}, {Fan}, \&
  {Wei}}]{2020ApJ...892...55J}
{Jiang}, J.-L., {Tang}, S.-P., {Wang}, Y.-Z., {Fan}, Y.-Z., \& {Wei}, D.-M.
  2020, \apj, 892, 55, \dodoi{10.3847/1538-4357/ab77cf}

\bibitem[{Karsch(2002)}]{Karsch:2001cy}
Karsch, F. 2002, Lect. Notes Phys., 583, 209, \dodoi{10.1007/3-540-45792-5_6}

\bibitem[{Kashyap {et~al.}(2022{\natexlab{a}})Kashyap, Dhani, \&
  Sathyaprakash}]{Kashyap:2022wzr}
Kashyap, R., Dhani, A., \& Sathyaprakash, B. 2022{\natexlab{a}}.
\newblock \doarXiv{2209.02757}

\bibitem[{Kashyap {et~al.}(2019)Kashyap, Raman, \& Ajith}]{Kashyap:2019ypm}
Kashyap, R., Raman, G., \& Ajith, P. 2019, Astrophys. J. Lett., 886, L19,
  \dodoi{10.3847/2041-8213/ab543f}

\bibitem[{Kashyap {et~al.}(2022{\natexlab{b}})}]{Kashyap:2021wzs}
Kashyap, R., {et~al.} 2022{\natexlab{b}}, Phys. Rev. D, 105, 103022,
  \dodoi{10.1103/PhysRevD.105.103022}

\bibitem[{Kasliwal {et~al.}(2019)Kasliwal, Kasen, Lau, Perley, Rosswog, Ofek,
  Hotokezaka, Chary, Sollerman, Goobar, \& Kaplan}]{KaKa2019}
Kasliwal, M.~M., Kasen, D., Lau, R.~M., {et~al.} 2019, MNRAS: Letters,
  \dodoi{10.1093/mnrasl/slz007}

\bibitem[{Keller {et~al.}(2022)Keller, Hebeler, \& Schwenk}]{Keller:2022crb}
Keller, J., Hebeler, K., \& Schwenk, A. 2022.
\newblock \doarXiv{2204.14016}

\bibitem[{Keller {et~al.}(2021)Keller, Wellenhofer, Hebeler, \&
  Schwenk}]{Keller:2020qhx}
Keller, J., Wellenhofer, C., Hebeler, K., \& Schwenk, A. 2021, Phys. Rev. C,
  103, 055806, \dodoi{10.1103/PhysRevC.103.055806}

\bibitem[{Kojo(2021)}]{Kojo:2021ugu}
Kojo, T. 2021, Phys. Rev. D, 104, 074005, \dodoi{10.1103/PhysRevD.104.074005}

\bibitem[{Komoltsev \& Kurkela(2022)}]{Komoltsev:2021jzg}
Komoltsev, O., \& Kurkela, A. 2022, Phys. Rev. Lett., 128, 202701,
  \dodoi{10.1103/PhysRevLett.128.202701}

\bibitem[{K\"oppel {et~al.}(2019)K\"oppel, Bovard, \&
  Rezzolla}]{Koppel:2019pys}
K\"oppel, S., Bovard, L., \& Rezzolla, L. 2019, Astrophys. J. Lett., 872, L16,
  \dodoi{10.3847/2041-8213/ab0210}

\bibitem[{{Kouvaris} \& {Tinyakov}(2010)}]{KouvarisTinyakov2010}
{Kouvaris}, C., \& {Tinyakov}, P. 2010, Phys. Rev. D, 82, 063531,
  \dodoi{10.1103/PhysRevD.82.063531}

\bibitem[{{Kramer} {et~al.}(2006){Kramer}, {Stairs}, {Manchester},
  {McLaughlin}, {Lyne}, {Ferdman}, {Burgay}, {Lorimer}, {Possenti}, {D'Amico},
  {Sarkissian}, {Hobbs}, {Reynolds}, {Freire}, \& {Camilo}}]{ksm+06}
{Kramer}, M., {Stairs}, I.~H., {Manchester}, R.~N., {et~al.} 2006, Science,
  314, 97, \dodoi{10.1126/science.1132305}

\bibitem[{Kurkela {et~al.}(2010)Kurkela, Romatschke, \&
  Vuorinen}]{Kurkela:2009gj}
Kurkela, A., Romatschke, P., \& Vuorinen, A. 2010, Phys. Rev. D, 81, 105021,
  \dodoi{10.1103/PhysRevD.81.105021}

\bibitem[{Kurkela \& Vuorinen(2016)}]{Kurkela:2016was}
Kurkela, A., \& Vuorinen, A. 2016, Phys. Rev. Lett., 117, 042501,
  \dodoi{10.1103/PhysRevLett.117.042501}

\bibitem[{{Kuulkers} {et~al.}(2003){Kuulkers}, {den Hartog}, {in't Zand},
  {Verbunt}, {Harris}, \& {Cocchi}}]{Kuulkers_etal_2003}
{Kuulkers}, E., {den Hartog}, P.~R., {in't Zand}, J.~J.~M., {et~al.} 2003,
  \aap, 399, 663, \dodoi{10.1051/0004-6361:20021781}

\bibitem[{{Lackey} \& {Wade}(2015)}]{2015PhRvD..91d3002L}
{Lackey}, B.~D., \& {Wade}, L. 2015, Phys. Rev. D, 91, 043002,
  \dodoi{10.1103/PhysRevD.91.043002}

\bibitem[{{Landry} {et~al.}(2020){Landry}, {Essick}, \&
  {Chatziioannou}}]{2020PhRvD.101l3007L}
{Landry}, P., {Essick}, R., \& {Chatziioannou}, K. 2020, Phys. Rev. D, 101,
  123007, \dodoi{10.1103/PhysRevD.101.123007}

\bibitem[{Landry \& Kumar(2018)}]{landry_constraints_2018}
Landry, P., \& Kumar, B. 2018, ApJL, 868, L22, \dodoi{10.3847/2041-8213/aaee76}

\bibitem[{{Lattimer}(2021)}]{Lattimer21}
{Lattimer}, J.~M. 2021, Annual Review of Nuclear and Particle Science, 71, 433,
  \dodoi{10.1146/annurev-nucl-102419-124827}

\bibitem[{{Lattimer} \& {Prakash}(2001{\natexlab{a}})}]{2001ApJ...550..426L}
{Lattimer}, J.~M., \& {Prakash}, M. 2001{\natexlab{a}}, \apj, 550, 426,
  \dodoi{10.1086/319702}

\bibitem[{{Lattimer} \& {Prakash}(2001{\natexlab{b}})}]{Lattimer01}
---. 2001{\natexlab{b}}, \apj, 550, 426, \dodoi{10.1086/319702}

\bibitem[{{Lattimer} \& {Prakash}(2005)}]{Lattimer05}
---. 2005, Physical Review Letters, 94, 111101,
  \dodoi{10.1103/PhysRevLett.94.111101}

\bibitem[{Lattimer \& Schramm(1974)}]{Lattimer:1974slx}
Lattimer, J.~M., \& Schramm, D.~N. 1974, Astrophys. J., 192, L145,
  \dodoi{10.1086/181612}

\bibitem[{Lattimer \& Schutz(2005)}]{lattimer_constraining_2005}
Lattimer, J.~M., \& Schutz, B.~F. 2005, ApJ, 629, 979, \dodoi{10.1086/431543}

\bibitem[{{Leung} {et~al.}(2011){Leung}, {Chu}, \& {Lin}}]{LeungChu2011}
{Leung}, S.~C., {Chu}, M.~C., \& {Lin}, L.~M. 2011, Phys. Rev. D, 84, 107301,
  \dodoi{10.1103/PhysRevD.84.107301}

\bibitem[{{Lewin} {et~al.}(1984){Lewin}, {Vacca}, \&
  {Basinska}}]{Lewin_etal_1984}
{Lewin}, W.~H.~G., {Vacca}, W.~D., \& {Basinska}, E.~M. 1984, \apjl, 277, L57,
  \dodoi{10.1086/184202}

\bibitem[{{Li} {et~al.}(2016){Li}, {Zhang}, {Zhang}, {Gao}, {Qi}, \&
  {Liu}}]{Li16}
{Li}, A., {Zhang}, B., {Zhang}, N.-B., {et~al.} 2016, \prd, 94, 083010,
  \dodoi{10.1103/PhysRevD.94.083010}

\bibitem[{{LIGO Scientific Collaboration} {et~al.}(2015){LIGO Scientific
  Collaboration}, {Aasi}, {Abbott}, {Abbott}, {Abbott}, {Abernathy}, {Ackley},
  {Adams}, {Adams}, {Addesso}, \& et~al.}]{LIGOScientificCollaborationAasi2015}
{LIGO Scientific Collaboration}, {Aasi}, J., {Abbott}, B.~P., {et~al.} 2015,
  Classical and Quantum Gravity, 32, 074001,
  \dodoi{10.1088/0264-9381/32/7/074001}

\bibitem[{Lim {et~al.}(2019)Lim, Holt, \& Stahulak}]{lim_predicting_2019}
Lim, Y., Holt, J.~W., \& Stahulak, R.~J. 2019, Phys. Rev. C, 100, 035802,
  \dodoi{10.1103/PhysRevC.100.035802}

\bibitem[{{Link} {et~al.}(1999){Link}, {Epstein}, \& {Lattimer}}]{linketal99}
{Link}, B., {Epstein}, R.~I., \& {Lattimer}, J.~M. 1999, \prl, 83, 3362,
  \dodoi{10.1103/PhysRevLett.83.3362}

\bibitem[{Lipunov {et~al.}(2017)}]{Lipunov:2017dwd}
Lipunov, V., {et~al.} 2017, Astrophys. J. Lett., 850, L1,
  \dodoi{10.3847/2041-8213/aa92c0}

\bibitem[{{Lo} {et~al.}(2013){Lo}, {Miller}, {Bhattacharyya}, \& {Lamb}}]{Lo13}
{Lo}, K.~H., {Miller}, M.~C., {Bhattacharyya}, S., \& {Lamb}, F.~K. 2013, \apj,
  776, 19, \dodoi{10.1088/0004-637X/776/1/19}

\bibitem[{{Loeb}(2003)}]{Loeb_2003}
{Loeb}, A. 2003, Physical Review Letters, 91, 071103,
  \dodoi{10.1103/PhysRevLett.91.071103}

\bibitem[{Manchester(2017)}]{wh16}
Manchester, R.~N. 2017, J. Astrophys. Astron., 38, 42,
  \dodoi{10.1007/s12036-017-9469-2}

\bibitem[{Margetis {et~al.}(2000)Margetis, Safarik, \&
  Villalobos~Baillie}]{Margetis:2000sv}
Margetis, S., Safarik, K., \& Villalobos~Baillie, O. 2000, Ann. Rev. Nucl.
  Part. Sci., 50, 299, \dodoi{10.1146/annurev.nucl.50.1.299}

\bibitem[{Martynov {et~al.}(2019)Martynov, Miao, Yang, Hernandez~Vivanco,
  Thrane, Smith, Lasky, East, Adhikari, Bauswein, Brooks, Chen, Corbitt,
  Corbitt, Grote, Levin, Zhao, \& Vecchio}]{Martynov2019}
Martynov, D., Miao, H., Yang, H., {et~al.} 2019, Phys. Rev. D, 99, 102004.
\newblock \doarXiv{1901.03885}

\bibitem[{McLerran \& Reddy(2019)}]{McLerran:2018hbz}
McLerran, L., \& Reddy, S. 2019, Phys. Rev. Lett., 122, 122701,
  \dodoi{10.1103/PhysRevLett.122.122701}

\bibitem[{Messenger {et~al.}(2014)Messenger, Takami, Gossan, Rezzolla, \&
  Sathyaprakash}]{Messenger:2013fya}
Messenger, C., Takami, K., Gossan, S., Rezzolla, L., \& Sathyaprakash, B.~S.
  2014, Phys. Rev. X, 4, 041004, \dodoi{10.1103/PhysRevX.4.041004}

\bibitem[{Miller {et~al.}(2019)}]{Miller:2019cac}
Miller, M., {et~al.} 2019, Astrophys. J. Lett., 887, L24,
  \dodoi{10.3847/2041-8213/ab50c5}

\bibitem[{{Miller}(2013)}]{miller13}
{Miller}, M.~C. 2013, ArXiv e-prints.
\newblock \doarXiv{1312.0029}

\bibitem[{{Miller} \& {Lamb}(1998)}]{Miller98}
{Miller}, M.~C., \& {Lamb}, F.~K. 1998, \apjl, 499, L37, \dodoi{10.1086/311335}

\bibitem[{{Miller} \& {Lamb}(2015)}]{MillerLamb15}
---. 2015, \apj, 808, 31, \dodoi{10.1088/0004-637X/808/1/31}

\bibitem[{{Miller} {et~al.}(2019){Miller}, {Lamb}, {Dittman}, B., Z., \&
  C.}]{miller19}
{Miller}, M.~C., {Lamb}, F.~K., {Dittman}, A.~J., {et~al.} 2019, \apjl, 887,
  L24, \dodoi{10.3847/2041-8213/ab50c5}

\bibitem[{{Miller} {et~al.}(2021){Miller}, {Lamb}, {Dittmann}, {Bogdanov},
  {Arzoumanian}, {Gendreau}, {Guillot}, {Ho}, {Lattimer}, {Loewenstein},
  {Morsink}, {Ray}, {Wolff}, {Baker}, {Cazeau}, {Manthripragada}, {Markwardt},
  {Okajima}, {Pollard}, {Cognard}, {Cromartie}, {Fonseca}, {Guillemot}, {Kerr},
  {Parthasarathy}, {Pennucci}, {Ransom}, \& {Stairs}}]{miller21}
{Miller}, M.~C., {Lamb}, F.~K., {Dittmann}, A.~J., {et~al.} 2021, \apjl, 918,
  L28, \dodoi{10.3847/2041-8213/ac089b}

\bibitem[{Montana {et~al.}(2019)Montana, Tolos, Hanauske, \&
  Rezzolla}]{Montana:2018bkb}
Montana, G., Tolos, L., Hanauske, M., \& Rezzolla, L. 2019, Phys. Rev. D, 99,
  103009, \dodoi{10.1103/PhysRevD.99.103009}

\bibitem[{Morrison {et~al.}(2004)Morrison, Baumgarte, Shapiro, \&
  Pandharipande}]{morrison_moment_2004}
Morrison, I.~A., Baumgarte, T.~W., Shapiro, S.~L., \& Pandharipande, V.~R.
  2004, ApJ, 617, L135, \dodoi{10.1086/427235}

\bibitem[{{Morsink} {et~al.}(2007){Morsink}, {Leahy}, {Cadeau}, \&
  {Braga}}]{Morsink07}
{Morsink}, S.~M., {Leahy}, D.~A., {Cadeau}, C., \& {Braga}, J. 2007, \apj, 663,
  1244, \dodoi{10.1086/518648}

\bibitem[{{Most} {et~al.}(2020){Most}, {Jens Papenfort}, {Dexheimer},
  {Hanauske}, {Stoecker}, \& {Rezzolla}}]{2020EPJA...56...59M}
{Most}, E.~R., {Jens Papenfort}, L., {Dexheimer}, V., {et~al.} 2020, European
  Physical Journal A, 56, 59, \dodoi{10.1140/epja/s10050-020-00073-4}

\bibitem[{{Most} {et~al.}(2019){Most}, {Papenfort}, {Dexheimer}, {Hanauske},
  {Schramm}, {St{\"o}cker}, \& {Rezzolla}}]{2019PhRvL.122f1101M}
{Most}, E.~R., {Papenfort}, L.~J., {Dexheimer}, V., {et~al.} 2019, Phys. Rev.
  Let., 122, 061101, \dodoi{10.1103/PhysRevLett.122.061101}

\bibitem[{Most {et~al.}(2018)Most, Weih, Rezzolla, \&
  Schaffner-Bielich}]{Most2018}
Most, E.~R., Weih, L.~R., Rezzolla, L., \& Schaffner-Bielich, J. 2018, Phys.
  Rev. Let., 120, 261103, \dodoi{10.1103/PhysRevLett.120.261103}

\bibitem[{{Motch} {et~al.}(2013){Motch}, {Wilms}, {Barret}, {Becker},
  {Bogdanov}, {Boirin}, {Corbel}, {Cackett}, {Campana}, {de Martino}, {Haberl},
  {in't Zand}, {M{\'e}ndez}, {Mignani}, {Miller}, {Orio}, {Psaltis}, {Rea},
  {Rodriguez}, {Rozanska}, {Schwope}, {Steiner}, {Webb}, {Zampieri}, \&
  {Zane}}]{2013arXiv1306.2334M}
{Motch}, C., {Wilms}, J., {Barret}, D., {et~al.} 2013, ArXiv e-prints.
\newblock \doarXiv{1306.2334}

\bibitem[{{Mushotzky}(2018)}]{2018SPIE10699E..29M}
{Mushotzky}, R. 2018, in Society of Photo-Optical Instrumentation Engineers
  (SPIE) Conference Series, Vol. 10699, Space Telescopes and Instrumentation
  2018: Ultraviolet to Gamma Ray, 1069929, \dodoi{10.1117/12.2310003}

\bibitem[{Nair \& Yunes(2020)}]{nair_improved_2020}
Nair, R., \& Yunes, N. 2020, Phys. Rev. D, 101, 104011,
  \dodoi{10.1103/PhysRevD.101.104011}

\bibitem[{{Nandra} {et~al.}(2013){Nandra}, {Barret}, {Barcons}, {Fabian}, {den
  Herder}, {Piro}, {Watson}, {Adami}, {Aird}, {Afonso}, \&
  et~al.}]{2013arXiv1306.2307N}
{Nandra}, K., {Barret}, D., {Barcons}, X., {et~al.} 2013, arXiv e-prints.
\newblock \doarXiv{1306.2307}

\bibitem[{{N{\"a}ttil{\"a}} {et~al.}(2017){N{\"a}ttil{\"a}}, {Miller},
  {Steiner}, {Kajava}, {Suleimanov}, \& {Poutanen}}]{Nattila17}
{N{\"a}ttil{\"a}}, J., {Miller}, M.~C., {Steiner}, A.~W., {et~al.} 2017, \aap,
  608, A31, \dodoi{10.1051/0004-6361/201731082}

\bibitem[{{N{\"a}ttil{\"a}} \& {Pihajoki}(2018)}]{Nattila18}
{N{\"a}ttil{\"a}}, J., \& {Pihajoki}, P. 2018, \aap, 615, A50,
  \dodoi{10.1051/0004-6361/201630261}

\bibitem[{Negele \& Vautherin(1973)}]{Negele:1971vb}
Negele, J.~W., \& Vautherin, D. 1973, Nucl. Phys. A, 207, 298,
  \dodoi{10.1016/0375-9474(73)90349-7}

\bibitem[{{Nelson} {et~al.}(2019){Nelson}, {Reddy}, \&
  {Zhou}}]{NelsonReddy2019}
{Nelson}, A.~E., {Reddy}, S., \& {Zhou}, D. 2019, J. Cosmol. Astropart. Phys.,
  2019, 012, \dodoi{10.1088/1475-7516/2019/07/012}

\bibitem[{Oppenheimer \& Volkoff(1939)}]{Oppenheimer:1939ne}
Oppenheimer, J.~R., \& Volkoff, G.~M. 1939, Phys. Rev., 55, 374,
  \dodoi{10.1103/PhysRev.55.374}

\bibitem[{{{\"O}zel}(2013)}]{ozel13}
{{\"O}zel}, F. 2013, Reports on Progress in Physics, 76, 016901,
  \dodoi{10.1088/0034-4885/76/1/016901}

\bibitem[{{{\"O}zel} {et~al.}(2010){{\"O}zel}, {Baym}, \&
  {G{\"u}ver}}]{Ozel_etal_2010}
{{\"O}zel}, F., {Baym}, G., \& {G{\"u}ver}, T. 2010, \prd, 82, 101301,
  \dodoi{10.1103/PhysRevD.82.101301}

\bibitem[{{{\"O}zel} \& {Freire}(2016)}]{Ozel_Freire_2016}
{{\"O}zel}, F., \& {Freire}, P. 2016, \araa, 54, 401,
  \dodoi{10.1146/annurev-astro-081915-023322}

\bibitem[{{{\"O}zel} {et~al.}(2009){{\"O}zel}, {G{\"u}ver}, \&
  {Psaltis}}]{Ozel_etal_2009}
{{\"O}zel}, F., {G{\"u}ver}, T., \& {Psaltis}, D. 2009, \apj, 693, 1775,
  \dodoi{10.1088/0004-637X/693/2/1775}

\bibitem[{{{\"O}zel} \& {Psaltis}(2003)}]{Ozel_Psaltis_2003}
{{\"O}zel}, F., \& {Psaltis}, D. 2003, \apjl, 582, L31, \dodoi{10.1086/346197}

\bibitem[{{{\"O}zel} \& {Psaltis}(2009)}]{Ozel09}
---. 2009, \prd, 80, 103003, \dodoi{10.1103/PhysRevD.80.103003}

\bibitem[{{Paerels}(1997)}]{Paerels_1997}
{Paerels}, F. 1997, \apjl, 476, L47, \dodoi{10.1086/310485}

\bibitem[{{Page} {et~al.}(2004){Page}, {Lattimer}, {Prakash}, \&
  {Steiner}}]{pagetal04}
{Page}, D., {Lattimer}, J.~M., {Prakash}, M., \& {Steiner}, A.~W. 2004, \apjs,
  155, 623, \dodoi{10.1086/424844}

\bibitem[{{Page} {et~al.}(2011){Page}, {Prakash}, {Lattimer}, \&
  {Steiner}}]{pagetal11}
{Page}, D., {Prakash}, M., {Lattimer}, J.~M., \& {Steiner}, A.~W. 2011, \prl,
  106, 081101, \dodoi{10.1103/PhysRevLett.106.081101}

\bibitem[{{Palatchi} \& {The CREX Collaboration
  Team}(2021)}]{2021APS..DNP.EA003P}
{Palatchi}, C., \& {The CREX Collaboration Team}. 2021, in APS Meeting
  Abstracts, Vol. 2021, APS Division of Nuclear Physics Meeting Abstracts,
  EA.003

\bibitem[{Pang {et~al.}(2020)Pang, Dietrich, Tews, \& Van
  Den~Broeck}]{Pang:2020ilf}
Pang, P.~T., Dietrich, T., Tews, I., \& Van Den~Broeck, C. 2020, arXiv
  e-prints, arXiv:2006.14936.
\newblock \doarXiv{2006.14936}

\bibitem[{{Panotopoulos} \& {Lopes}(2017)}]{PanotopoulosLopes2017}
{Panotopoulos}, G., \& {Lopes}, I. 2017, Phys. Rev. D, 96, 083004,
  \dodoi{10.1103/PhysRevD.96.083004}

\bibitem[{{Pechenick} {et~al.}(1983){Pechenick}, {Ftaclas}, \&
  {Cohen}}]{Pechenick83}
{Pechenick}, K.~R., {Ftaclas}, C., \& {Cohen}, J.~M. 1983, \apj, 274, 846,
  \dodoi{10.1086/161498}

\bibitem[{{Potekhin}(2014)}]{potekhin14}
{Potekhin}, A.~Y. 2014, Physics Uspekhi, 57, 735,
  \dodoi{10.3367/UFNe.0184.201408a.0793}

\bibitem[{{Poutanen} \& {Beloborodov}(2006)}]{Poutanen06}
{Poutanen}, J., \& {Beloborodov}, A.~M. 2006, \mnras, 373, 836,
  \dodoi{10.1111/j.1365-2966.2006.11088.x}

\bibitem[{{Poutanen} \& {Gierli{\'n}ski}(2003)}]{Poutanen03}
{Poutanen}, J., \& {Gierli{\'n}ski}, M. 2003, \mnras, 343, 1301,
  \dodoi{10.1046/j.1365-8711.2003.06773.x}

\bibitem[{{Psaltis} {et~al.}(2014){Psaltis}, {{\"O}zel}, \&
  {Chakrabarty}}]{Psaltis14}
{Psaltis}, D., {{\"O}zel}, F., \& {Chakrabarty}, D. 2014, \apj, 787, 136,
  \dodoi{10.1088/0004-637X/787/2/136}

\bibitem[{Punturo {et~al.}(2010)}]{Punturo:2010zz}
Punturo, M., {et~al.} 2010, Class. Quant. Grav., 27, 194002,
  \dodoi{10.1088/0264-9381/27/19/194002}

\bibitem[{{Quddus} {et~al.}(2020){Quddus}, {Panotopoulos}, {Kumar}, {Ahmad}, \&
  {Patra}}]{QuddusPanotopoulos2020}
{Quddus}, A., {Panotopoulos}, G., {Kumar}, B., {Ahmad}, S., \& {Patra}, S.~K.
  2020, Journal of Physics G Nuclear Physics, 47, 095202,
  \dodoi{10.1088/1361-6471/ab9d36}

\bibitem[{{Raaijmakers} {et~al.}(2018){Raaijmakers}, {Riley}, \&
  {Watts}}]{Raaijmakers18}
{Raaijmakers}, G., {Riley}, T.~E., \& {Watts}, A.~L. 2018, \mnras, 478, 2177,
  \dodoi{10.1093/mnras/sty1052}

\bibitem[{{Raaijmakers} {et~al.}(2020){Raaijmakers}, {Greif}, {Riley},
  {Hinderer}, {Hebeler}, {Schwenk}, {Watts}, {Nissanke}, {Guillot}, {Lattimer},
  \& {Ludlam}}]{raaijmakers20}
{Raaijmakers}, G., {Greif}, S.~K., {Riley}, T.~E., {et~al.} 2020, \apjl, 893,
  L21, \dodoi{10.3847/2041-8213/ab822f}

\bibitem[{Raaijmakers {et~al.}(2021)Raaijmakers, Greif, Hebeler, Hinderer,
  Nissanke, Schwenk, Riley, Watts, Lattimer, \& Ho}]{Raaijmakers:2021uju}
Raaijmakers, G., Greif, S.~K., Hebeler, K., {et~al.} 2021, Astrophys. J. Lett.,
  918, L29, \dodoi{10.3847/2041-8213/ac089a}

\bibitem[{{Radice} {et~al.}(2017){Radice}, {Bernuzzi}, {Del Pozzo}, {Roberts},
  \& {Ott}}]{2017ApJ...842L..10R}
{Radice}, D., {Bernuzzi}, S., {Del Pozzo}, W., {Roberts}, L.~F., \& {Ott},
  C.~D. 2017, Astrophys. J. Let., 842, L10, \dodoi{10.3847/2041-8213/aa775f}

\bibitem[{{Radice} {et~al.}(2020){Radice}, {Bernuzzi}, \&
  {Perego}}]{2020arXiv200203863R}
{Radice}, D., {Bernuzzi}, S., \& {Perego}, A. 2020, arXiv e-prints,
  arXiv:2002.03863.
\newblock \doarXiv{2002.03863}

\bibitem[{Radice \& Dai(2018)}]{Radice2018a}
Radice, D., \& Dai, L. 2018, Eur. Phys. J. A, 55, 50.
\newblock \doarXiv{1810.12917}

\bibitem[{{Raithel}(2019)}]{2019EPJA...55...80R}
{Raithel}, C.~A. 2019, European Physical Journal A, 55, 80,
  \dodoi{10.1140/epja/i2019-12759-5}

\bibitem[{Raithel {et~al.}(2016)Raithel, Özel, \&
  Psaltis}]{raithel_model-independent_2016}
Raithel, C.~A., Özel, F., \& Psaltis, D. 2016, Phys. Rev. C, 93, 032801,
  \dodoi{10.1103/PhysRevC.93.032801}

\bibitem[{{Ransom} {et~al.}(2014){Ransom}, {Stairs}, {Archibald}, {Hessels},
  {Kaplan}, {van Kerkwijk}, {Boyles}, {Deller}, {Chatterjee},
  {Schechtman-Rook}, {Berndsen}, {Lynch}, {Lorimer}, {Karako-Argaman}, {Kaspi},
  {Kondratiev}, {McLaughlin}, {van Leeuwen}, {Rosen}, {Roberts}, \&
  {Stovall}}]{rsa+14}
{Ransom}, S.~M., {Stairs}, I.~H., {Archibald}, A.~M., {et~al.} 2014, Nature,
  505, 520, \dodoi{10.1038/nature12917}

\bibitem[{{Ray} {et~al.}(2019){Ray}, {Arzoumanian}, {Ballantyne}, {Bozzo},
  {Brandt}, {Brenneman}, {Chakrabarty}, {Christophersen}, {DeRosa}, {Feroci},
  {Gendreau}, {Goldstein}, {Hartmann}, {Hernanz}, {Jenke}, {Kara}, {Maccarone},
  {McDonald}, {Nowak}, {Phlips}, {Remillard}, {Stevens}, {Tomsick}, {Watts},
  {Wilson-Hodge}, {Wood}, \& {Zane}}]{2019arXiv190303035R}
{Ray}, P.~S., {Arzoumanian}, Z., {Ballantyne}, D., {et~al.} 2019, arXiv
  e-prints.
\newblock \doarXiv{1903.03035}

\bibitem[{{Read} {et~al.}(2009){Read}, {Lackey}, {Owen}, \&
  {Friedman}}]{Read09a}
{Read}, J.~S., {Lackey}, B.~D., {Owen}, B.~J., \& {Friedman}, J.~L. 2009, \prd,
  79, 124032, \dodoi{10.1103/PhysRevD.79.124032}

\bibitem[{Reitze {et~al.}(2019)}]{Reitze:2019iox}
Reitze, D., {et~al.} 2019, Bull. Am. Astron. Soc., 51, 035.
\newblock \doarXiv{1907.04833}

\bibitem[{{Riles}(2022)}]{2022arXiv220606447R}
{Riles}, K. 2022, arXiv e-prints, arXiv:2206.06447.
\newblock \doarXiv{2206.06447}

\bibitem[{{Riley} {et~al.}(2018){Riley}, {Raaijmakers}, \& {Watts}}]{Riley18}
{Riley}, T.~E., {Raaijmakers}, G., \& {Watts}, A.~L. 2018, \mnras, 478, 1093,
  \dodoi{10.1093/mnras/sty1051}

\bibitem[{Riley {et~al.}(2019)}]{Riley:2019yda}
Riley, T.~E., {et~al.} 2019, Astrophys. J. Lett., 887, L21,
  \dodoi{10.3847/2041-8213/ab481c}

\bibitem[{{Riley} {et~al.}(2021){Riley}, {Watts}, {Ray}, {Bogdanov}, {Guillot},
  {Morsink}, {Bilous}, {Arzoumanian}, {Choudhury}, {Deneva}, {Gendreau},
  {Harding}, {Ho}, {Lattimer}, {Loewenstein}, {Ludlam}, {Markwardt}, {Okajima},
  {Prescod-Weinstein}, {Remillard}, {Wolff}, {Fonseca}, {Cromartie}, {Kerr},
  {Pennucci}, {Parthasarathy}, {Ransom}, {Stairs}, {Guillemot}, \&
  {Cognard}}]{riley21}
{Riley}, T.~E., {Watts}, A.~L., {Ray}, P.~S., {et~al.} 2021, \apjl, 918, L27,
  \dodoi{10.3847/2041-8213/ac0a81}

\bibitem[{Romatschke \& Romatschke(2019)}]{Romatschke:2017ejr}
Romatschke, P., \& Romatschke, U. 2019, {Relativistic Fluid Dynamics In and Out
  of Equilibrium}, Cambridge Monographs on Mathematical Physics (Cambridge
  University Press), \dodoi{10.1017/9781108651998}

\bibitem[{{Rutherford} {et~al.}(2022){Rutherford}, {Raaijmakers},
  {Prescod-Weinstein}, \& {Watts}}]{2022arXiv220803282R}
{Rutherford}, N., {Raaijmakers}, G., {Prescod-Weinstein}, C., \& {Watts}, A.
  2022, arXiv e-prints, arXiv:2208.03282.
\newblock \doarXiv{2208.03282}

\bibitem[{{Rutledge} {et~al.}(2002){Rutledge}, {Bildsten}, {Brown}, {Pavlov},
  \& {Zavlin}}]{Rutledge_etal_2002}
{Rutledge}, R.~E., {Bildsten}, L., {Brown}, E.~F., {Pavlov}, G.~G., \&
  {Zavlin}, V.~E. 2002, \apj, 578, 405, \dodoi{10.1086/342306}

\bibitem[{Samajdar \& Dietrich(2020)}]{Samajdar:2020xrd}
Samajdar, A., \& Dietrich, T. 2020, Phys. Rev. D, 101, 124014,
  \dodoi{10.1103/PhysRevD.101.124014}

\bibitem[{Savchenko {et~al.}(2017)Savchenko, Ferrigno, Kuulkers, Bazzano,
  Bozzo, Brandt, Chenevez, Courvoisier, Diehl, Domingo, Hanlon, Jourdain, von
  Kienlin, Laurent, Lebrun, Lutovinov, Martin-Carrillo, Mereghetti, Natalucci,
  Rodi, Roques, Sunyaev, \& Ubertini}]{AbEA2017e}
Savchenko, V., Ferrigno, C., Kuulkers, E., {et~al.} 2017, The Astrophysical
  Journal Letters, 848, L15.
\newblock \url{http://stacks.iop.org/2041-8205/848/i=2/a=L15}

\bibitem[{Sennett {et~al.}(2017)Sennett, Hinderer, Steinhoff, Buonanno, \&
  Ossokine}]{Sennett:2017etc}
Sennett, N., Hinderer, T., Steinhoff, J., Buonanno, A., \& Ossokine, S. 2017,
  Phys. Rev. D, 96, 024002, \dodoi{10.1103/PhysRevD.96.024002}

\bibitem[{{Shapiro}(1964)}]{sha64}
{Shapiro}, I.~I. 1964, Physical Review Letters, 13, 789,
  \dodoi{10.1103/PhysRevLett.13.789}

\bibitem[{Shibata {et~al.}(2005)Shibata, Taniguchi, \&
  Ury{\=u}}]{Shibata:2005ss}
Shibata, M., Taniguchi, K., \& Ury{\=u}, K. 2005, Phys. Rev. D, 71, 084021,
  \dodoi{10.1103/PhysRevD.71.084021}

\bibitem[{{Shternin} {et~al.}(2021){Shternin}, {Ofengeim}, {Ho}, {Heinke},
  {Wijngaarden}, \& {Patnaude}}]{shterninetal21}
{Shternin}, P.~S., {Ofengeim}, D.~D., {Ho}, W. C.~G., {et~al.} 2021, \mnras,
  506, 709, \dodoi{10.1093/mnras/stab1695}

\bibitem[{{Shternin} {et~al.}(2011){Shternin}, {Yakovlev}, {Heinke}, {Ho}, \&
  {Patnaude}}]{shterninetal11}
{Shternin}, P.~S., {Yakovlev}, D.~G., {Heinke}, C.~O., {Ho}, W. C.~G., \&
  {Patnaude}, D.~J. 2011, \mnras, 412, L108,
  \dodoi{10.1111/j.1745-3933.2011.01015.x}

\bibitem[{Silva {et~al.}(2020)Silva, Holgado, Cárdenas-Avendaño, \&
  Yunes}]{silva_astrophysical_2020}
Silva, H.~O., Holgado, A.~M., Cárdenas-Avendaño, A., \& Yunes, N. 2020,
  arXiv:2004.01253.
\newblock \url{http://arxiv.org/abs/2004.01253}

\bibitem[{Singer {et~al.}(2014)}]{Singer:2014qca}
Singer, L.~P., {et~al.} 2014, Astrophys. J., 795, 105,
  \dodoi{10.1088/0004-637X/795/2/105}

\bibitem[{Soares-Santos {et~al.}(2017)}]{Soares-Santos:2017lru}
Soares-Santos, M., {et~al.} 2017, Astrophys. J. Lett., 848, L16,
  \dodoi{10.3847/2041-8213/aa9059}

\bibitem[{Soultanis {et~al.}(2022)Soultanis, Bauswein, \&
  Stergioulas}]{Soultanis:2021oia}
Soultanis, T., Bauswein, A., \& Stergioulas, N. 2022, Phys. Rev. D, 105,
  043020, \dodoi{10.1103/PhysRevD.105.043020}

\bibitem[{Srivastava {et~al.}(2022)Srivastava, Davis, Kuns, Landry, Ballmer,
  Evans, Hall, Read, \& Sathyaprakash}]{Srivastava:2022slt}
Srivastava, V., Davis, D., Kuns, K., {et~al.} 2022, Astrophys. J., 931, 22,
  \dodoi{10.3847/1538-4357/ac5f04}

\bibitem[{{Stairs} {et~al.}(2005){Stairs}, {Faulkner}, {Lyne}, {Kramer},
  {Lorimer}, {McLaughlin}, {Manchester}, {Hobbs}, {Camilo}, {Possenti},
  {Burgay}, {D'Amico}, {Freire}, \& {Gregory}}]{sfl+05}
{Stairs}, I.~H., {Faulkner}, A.~J., {Lyne}, A.~G., {et~al.} 2005, ApJ, 632,
  1060, \dodoi{10.1086/432526}

\bibitem[{{Steiner} {et~al.}(2018){Steiner}, {Heinke}, {Bogdanov}, {Li}, {Ho},
  {Bahramian}, \& {Han}}]{Steiner_etal_2018}
{Steiner}, A.~W., {Heinke}, C.~O., {Bogdanov}, S., {et~al.} 2018, \mnras, 476,
  421, \dodoi{10.1093/mnras/sty215}

\bibitem[{{Steiner} {et~al.}(2010){Steiner}, {Lattimer}, \&
  {Brown}}]{Steiner_etal_2010}
{Steiner}, A.~W., {Lattimer}, J.~M., \& {Brown}, E.~F. 2010, \apj, 722, 33,
  \dodoi{10.1088/0004-637X/722/1/33}

\bibitem[{{Steiner} {et~al.}(2005){Steiner}, {Prakash}, {Lattimer}, \&
  {Ellis}}]{sple05}
{Steiner}, A.~W., {Prakash}, M., {Lattimer}, J.~M., \& {Ellis}, P.~J. 2005,
  Phys. Rep., 411, 325, \dodoi{10.1016/j.physrep.2005.02.004}

\bibitem[{Takhistov(2018)}]{Takhistov:2017bpt}
Takhistov, V. 2018, Phys. Lett. B, 782, 77,
  \dodoi{10.1016/j.physletb.2018.05.026}

\bibitem[{Tanvir {et~al.}(2017)}]{Tanvir:2017pws}
Tanvir, N., {et~al.} 2017, Astrophys. J. Lett., 848, L27,
  \dodoi{10.3847/2041-8213/aa90b6}

\bibitem[{{Tawara} {et~al.}(1984){Tawara}, {Kii}, {Hayakawa}, {Kunieda},
  {Masai}, {Nagase}, {Inoue}, {Koyama}, {Makino}, {Makishima}, {Matsuoka},
  {Murakami}, {Oda}, {Ogawara}, {Ohashi}, {Shibazaki}, {Tanaka}, {Miyamoto},
  {Tsunemi}, {Yamashita}, \& {Kondo}}]{Tawara_etal_1984}
{Tawara}, Y., {Kii}, T., {Hayakawa}, S., {et~al.} 1984, \apjl, 276, L41,
  \dodoi{10.1086/184184}

\bibitem[{Taylor \& Weisberg(1982)}]{taylor_new_1982}
Taylor, J.~H., \& Weisberg, J.~M. 1982, ApJ, 253, 908, \dodoi{10.1086/159690}

\bibitem[{Taylor \& Gair(2012)}]{PhysRevD.86.023502}
Taylor, S.~R., \& Gair, J.~R. 2012, Phys. Rev. D, 86, 023502,
  \dodoi{10.1103/PhysRevD.86.023502}

\bibitem[{{Tews} {et~al.}(2019){Tews}, {Margueron}, \&
  {Reddy}}]{2019EPJA...55...97T}
{Tews}, I., {Margueron}, J., \& {Reddy}, S. 2019, European Physical Journal A,
  55, 97, \dodoi{10.1140/epja/i2019-12774-6}

\bibitem[{{The LIGO Scientific Collaboration} {et~al.}(2020){The LIGO
  Scientific Collaboration}, {the Virgo Collaboration}, Abbott, Abbott, Abbott,
  Abraham, Acernese, Ackley, Adams, Adhikari, Adya, Affeldt, \&
  et~al.}]{lvc2020}
{The LIGO Scientific Collaboration}, {the Virgo Collaboration}, Abbott, B.~P.,
  {et~al.} 2020, Astrophys. J. Let., 892, L3, \dodoi{10.3847/2041-8213/ab75f5}

\bibitem[{{The LIGO Scientific Collaboration and Virgo
  Collaboration}(2017)}]{AbEA2017g}
{The LIGO Scientific Collaboration and Virgo Collaboration}. 2017, The
  Astrophysical Journal Letters, 850, L39.
\newblock \url{http://stacks.iop.org/2041-8205/850/i=2/a=L39}

\bibitem[{{The Lynx Team}(2018)}]{2018arXiv180909642T}
{The Lynx Team}. 2018, arXiv e-prints, arXiv:1809.09642.
\newblock \doarXiv{1809.09642}

\bibitem[{Thiel {et~al.}(2019)Thiel, Sfienti, Piekarewicz, Horowitz, \&
  Vanderhaeghen}]{Thiel:2019tkm}
Thiel, M., Sfienti, C., Piekarewicz, J., Horowitz, C.~J., \& Vanderhaeghen, M.
  2019, J. Phys. G, 46, 093003, \dodoi{10.1088/1361-6471/ab2c6d}

\bibitem[{Tolman(1939)}]{Tolman:1939jz}
Tolman, R.~C. 1939, Phys. Rev., 55, 364, \dodoi{10.1103/PhysRev.55.364}

\bibitem[{Tootle {et~al.}(2021)Tootle, Papenfort, Most, \&
  Rezzolla}]{Tootle:2021umi}
Tootle, S.~D., Papenfort, L.~J., Most, E.~R., \& Rezzolla, L. 2021, Astrophys.
  J. Lett., 922, L19, \dodoi{10.3847/2041-8213/ac350d}

\bibitem[{Torres-Rivas {et~al.}(2019)Torres-Rivas, Chatziioannou, Bauswein, \&
  Clark}]{Torres-Rivas2019}
Torres-Rivas, A., Chatziioannou, K., Bauswein, A., \& Clark, J.~A. 2019, Phys.
  Rev. D, 99, 044014, \dodoi{10.1103/PhysRevD.99.044014}

\bibitem[{Tsang {et~al.}(2019)Tsang, Dietrich, \& Van
  Den~Broeck}]{Tsang:2019esi}
Tsang, K.~W., Dietrich, T., \& Van Den~Broeck, C. 2019, Phys. Rev. D, 100,
  044047, \dodoi{10.1103/PhysRevD.100.044047}

\bibitem[{Valenti {et~al.}(2017)Valenti, Sand, Yang, Cappellaro, Tartaglia,
  Corsi, Jha, Reichart, Haislip, \& Kouprianov}]{Valenti:2017ngx}
Valenti, S., Sand, D.~J., Yang, S., {et~al.} 2017, Astrophys. J. Lett., 848,
  L24, \dodoi{10.3847/2041-8213/aa8edf}

\bibitem[{{Venkatraman Krishnan} {et~al.}(2020){Venkatraman Krishnan},
  {Bailes}, {van Straten}, {Wex}, {Freire}, {Keane}, {Tauris}, {Rosado},
  {Bhat}, {Flynn}, {Jameson}, \& {Os{\l}owski}}]{kbv+20}
{Venkatraman Krishnan}, V., {Bailes}, M., {van Straten}, W., {et~al.} 2020,
  Science, 367, 577, \dodoi{10.1126/science.aax7007}

\bibitem[{{Voisin} {et~al.}(2020){Voisin}, {Cognard}, {Freire}, {Wex},
  {Guillemot}, {Desvignes}, {Kramer}, \& {Theureau}}]{vcf+20}
{Voisin}, G., {Cognard}, I., {Freire}, P.~C.~C., {et~al.} 2020, Astron. \&
  Astrophy., 638, A24, \dodoi{10.1051/0004-6361/202038104}

\bibitem[{{Vretinaris} {et~al.}(2020){Vretinaris}, {Stergioulas}, \&
  {Bauswein}}]{2020PhRvD.101h4039V}
{Vretinaris}, S., {Stergioulas}, N., \& {Bauswein}, A. 2020, Phys. Rev. D, 101,
  084039, \dodoi{10.1103/PhysRevD.101.084039}

\bibitem[{{Watanabe} \& {Pethick}(2017)}]{watanabepethick17}
{Watanabe}, G., \& {Pethick}, C.~J. 2017, \prl, 119, 062701,
  \dodoi{10.1103/PhysRevLett.119.062701}

\bibitem[{Watson {et~al.}(2019)}]{Watson:2019xjv}
Watson, D., {et~al.} 2019, Nature, 574, 497, \dodoi{10.1038/s41586-019-1676-3}

\bibitem[{{Watts} {et~al.}(2016){Watts}, {Andersson}, {Chakrabarty}, {Feroci},
  {Hebeler}, {Israel}, {Lamb}, {Miller}, {Morsink}, {{\"O}zel}, {Patruno},
  {Poutanen}, {Psaltis}, {Schwenk}, {Steiner}, {Stella}, {Tolos}, \& {van der
  Klis}}]{Watts16}
{Watts}, A.~L., {Andersson}, N., {Chakrabarty}, D., {et~al.} 2016, Reviews of
  Modern Physics, 88, 021001, \dodoi{10.1103/RevModPhys.88.021001}

\bibitem[{{Weih} {et~al.}(2019){Weih}, {Hanauske}, \&
  {Rezzolla}}]{2019arXiv191209340W}
{Weih}, L.~R., {Hanauske}, M., \& {Rezzolla}, L. 2019, arXiv e-prints,
  arXiv:1912.09340.
\newblock \doarXiv{1912.09340}

\bibitem[{Weisberg {et~al.}(1981)Weisberg, Taylor, \&
  Fowler}]{weisberg_gravitational_1981}
Weisberg, J.~M., Taylor, J.~H., \& Fowler, L.~A. 1981, Scientific American,
  245, 74, \dodoi{10.1038/scientificamerican1081-74}

\bibitem[{Wuchner {et~al.}(2022)Wuchner, Landry, \& Read}]{WuchnerPrep}
Wuchner, E., Landry, P., \& Read, J. 2022, Recovering masses and tides from
  neutron star mergers \emph{in preparation}.

\bibitem[{Wysocki {et~al.}(2020)Wysocki, O'Shaughnessy, Wade, \&
  Lange}]{Wysocki:2020myz}
Wysocki, D., O'Shaughnessy, R., Wade, L., \& Lange, J. 2020, arXiv e-prints,
  arXiv:2001.01747.
\newblock \doarXiv{2001.01747}

\bibitem[{Yagi \& Yunes(2013)}]{yagi_i-love-q_2013}
Yagi, K., \& Yunes, N. 2013, Science, 341, 365, \dodoi{10.1126/science.1236462}

\bibitem[{Yang {et~al.}(2018)Yang, Paschalidis, Yagi, Lehner, Pretorius, \&
  Yunes}]{Yang2018}
Yang, H., Paschalidis, V., Yagi, K., {et~al.} 2018, Phys. Rev. D, 97, 024049,
  \dodoi{10.1103/PhysRevD.97.024049}

\bibitem[{Yunes \& Hughes(2010)}]{yunes_binary_2010}
Yunes, N., \& Hughes, S.~A. 2010, Phys. Rev. D, 82, 082002,
  \dodoi{10.1103/PhysRevD.82.082002}

\bibitem[{{Zdunik} \& {Haensel}(2013)}]{Zdunik13}
{Zdunik}, J.~L., \& {Haensel}, P. 2013, \aap, 551, A61,
  \dodoi{10.1051/0004-6361/201220697}

\bibitem[{{Zhao} \& {Lattimer}(2022)}]{Zhao-Lattimer2022}
{Zhao}, T., \& {Lattimer}, J. 2022

\bibitem[{{Zhu} {et~al.}(2019){Zhu}, {Desvignes}, {Wex}, {Caballero},
  {Champion}, {Demorest}, {Ellis}, {Janssen}, {Kramer}, {Krieger}, {Lentati},
  {Nice}, {Ransom}, {Stairs}, {Stappers}, {Verbiest}, {Arzoumanian}, {Bassa},
  {Burgay}, {Cognard}, {Crowter}, {Dolch}, {Ferdman}, {Fonseca}, {Gonzalez},
  {Graikou}, {Guillemot}, {Hessels}, {Jessner}, {Jones}, {Jones}, {Jordan},
  {Karuppusamy}, {Lam}, {Lazaridis}, {Lazarus}, {Lee}, {Levin}, {Liu}, {Lyne},
  {McKee}, {McLaughlin}, {Os{\l}owski}, {Pennucci}, {Perrodin}, {Possenti},
  {Sanidas}, {Shaifullah}, {Smits}, {Stovall}, {Swiggum}, {Theureau}, \&
  {Tiburzi}}]{zdw+19}
{Zhu}, W.~W., {Desvignes}, G., {Wex}, N., {et~al.} 2019, MNRAS, 482, 3249,
  \dodoi{10.1093/mnras/sty2905}

\end{thebibliography}

\end{document}